%% file: main.tex
\documentclass[12pt]{article}

\usepackage[left=1.25in,right=1.25in]{geometry}
\usepackage{setspace}
\usepackage[backend=biber, style=apa,maxcitenames=1, maxbibnames=1, minnames=1, maxnames=1]{biblatex}

\usepackage{pdflscape}  
\usepackage{graphicx}
\usepackage{subcaption}
\usepackage{amsmath}
\usepackage{dsfont}
\usepackage{longtable}
\usepackage{caption}
\usepackage{rotating}
\usepackage{booktabs,makecell,siunitx}
\usepackage{longtable,tabularx,ragged2e,array}
\usepackage{threeparttablex}
\usepackage{newpxtext,newpxmath}
\usepackage{authblk}

\usepackage[colorlinks=true, urlcolor=blue, linkcolor=red]{hyperref}
\title{\vspace{-1.5em}\textbf{Measuring Sentiment News with Transformer-Based Language Models}}

\author[1]{Maria Saveria Mavillonio}
\author[2]{Stefano Borgioli}
\author[1]{Caterina Giannetti\thanks{Corresponding author: Caterina Giannetti,
\href{mailto:caterina.giannetti@unipi.it}{caterina.giannetti@unipi.it}.
The authors acknowledge support from the project "Teaming-up with social
artificial agents", funded by the Program for Research Projects of National
Interest (PRIN), grant no. 2022ALBSWX.}}
\author[2]{Chiara Ongari}
\author[3]{Giampiero M. Gallo}

\affil[1]{\small University of Pisa} 
\affil[2]{\small European Central Bank} 
\affil[3]{\small New York University in Florence} 

\date{\small  \today}

\begin{document}
\maketitle

\begin{abstract}
\begin{spacing}{1}
Measuring sentiment from financial news is a central task in economics and finance, yet most existing indicators rely on dictionary-based approaches that infer sentiment from word counts and only partially capture context, negation, and semantic structure. This paper proposes a framework for constructing daily news mood indices using transformer-based language models and evaluates whether they better represent sentiment than dictionary-based alternatives. Using 143,755 financial news articles from Factiva, we classify sentiment at the sentence level with FinBERT and aggregate these predictions into article-level and daily sentiment measures through alternative normalization schemes. We compare the resulting indices with benchmark measures based on \cite{shapiro2022measuring} and \cite{barbaglia2025sentiment}. A central contribution is the validation of alternative sentiment measures against human judgments. We conducted an incentivized annotation exercise in which 444 participants evaluated a validation subsample of 588 financial news articles. Consensus ratings from independent human evaluations serve as an external benchmark for assessing the quality of automated sentiment measures. Across correlation, regression, and classification exercises, transformer-based measures show stronger agreement with human judgments than vocabulary-based alternatives and perform substantially better  in distinguishing positive, neutral, and negative articles. Overall, the results suggest that incorporating contextual information through transformer-based language models produces sentiment measures that more closely reflect human assessments of financial news.
\end{spacing}
\end{abstract}



\section{Introduction}
Timely information on the state of the economy is essential for both policymakers and market participants. However, core macroeconomic aggregates—such as GDP, employment, and inflation—are typically observed at relatively low frequency and are released with non-negligible publication lags. A range of approaches has been developed in the economic literature and in practice to address this informational time gap, including “nowcasting” techniques.
An alternative class of models exploits information available across heterogeneous media sources and formats. A central component of this research agenda is the construction of sentiment, or “mood,” indices derived from news data, designed to capture systematic variation in the tone of economic reporting at daily or higher frequencies.
For instance, large corpora of news articles may be transformed into high-frequency time series designed to proxy narratives, expectations, and “animal spirits”; these metrics are subsequently incorporated into forecasting and structural modelling frameworks.

In addition to traditional media, recent contributions increasingly exploit high-frequency digital text, including social media posts (e.g., Twitter/X), online search activity, and other forms of web-based content. These sources are particularly valuable, as they react very rapidly to economic developments and may capture dimensions of attention and sentiment that are not fully reflected in conventional news outlets \parencite{marcucci2024}.

A rather recent but already widely used benchmark in this stream of literature is \textcite{shapiro2022measuring}, who constructs a news-based measure designed to mimic survey indicators by combining textual information with a transparent aggregation strategy. Such dictionary- and co-occurrence-based estimation procedures, valued for their interpretability and scalability, have become standard tools in economics and finance. At the same time, these approaches share an important limitation: sentiment is inferred primarily from isolated tokens or local word co-occurrence statistics, rather than from the broader semantic context in which language is used.
As a result, they are less well equipped to capture subtleties in sentiment such as negation, compositional meaning, and context-dependent interpretations of tone. In economic news, however, polarity is often determined by syntactic structure and pragmatic context. For example, modal expressions (e.g., "could worsen") convey uncertainty that affects sentiment, while domain-specific usage can overturn interpretations: phrases such as "lower inflation" are typically associated with positive sentiment, whereas "lower growth" conveys a negative outlook. In fact, when sentiment is proxied through word counts, such contextual cues are often overlooked, leading to attenuation bias and systematic misclassification—particularly in instances where tonal variation is most economically informative. Consequently, recent research has advanced lexicon-based sentiment measurement through more rigorous domain-specific adaptation. For instance, \textcite{barbaglia2025sentiment} introduce an Economic Lexicon (EL) specifically tailored to economic texts, representing an improvement over general-purpose and finance/accounting dictionaries. The EL assigns continuous sentiment scores in the range [-1,1] to words based on human annotation within economic contexts, while excluding highly ambiguous terms. The lexicon is constructed by integrating large corpora of US and UK news with central-bank publications, isolating sentences pertaining to economic concepts, and applying dependency parsing to identify candidate modifiers that convey sentiment.


This paper revisits the construction of daily mood indices by replicating the core analysis of  \textcite{shapiro2022measuring} using a more context-sensitive scoring methodology based on transformer language models \parencite{devlin2018bert}. Specifically, we employ encoder architectures from the BERT family, fine-tuned for financial sentiment, to map text into sentence-level sentiment probabilities that account for full contextual information. Unlike traditional classifiers, this approach fundamentally alters how sentiment is measured: rather than relying on counts of positive and negative words, it infers sentiment from the meaning of each sentence in context. The methodology is designed to enhance measurement accuracy in settings where dictionary-based methods typically underperform—such as in the presence of negation, hedging, and complex semantic dependencies—while maintaining the transparent aggregation framework that underpins the practical utility of news-based sentiment indices in economic applications.

Our contributions are threefold. First, we construct daily news mood indices using transformer encoders within a transparent and reproducible aggregation framework, assessing how alternative treatments of neutrality, text length, and sentiment imbalance affect the resulting time series. Second, we compare the transformer-based indices with a Shapiro-style benchmark and examine whether deviations arise from sentence composition, publisher mix, and calendar effects. Third, we validate the alternative indices through a resource-intensive human annotation exercise, treating human judgments as the ground truth and conducting both in-sample and out-of-sample evaluations.

From the operational strategy point of view, we follow \textcite{shapiro2022measuring} sequential framework, but we modify the scoring stage by employing transformer-based language models. Each article is segmented into sentences, for which we obtain class probabilities across positive, negative, and neutral categories. These sentence-level predictions are then aggregated to the article level using a variety of normalization schemes that balance robustness to neutral content against sensitivity to polarized language. Subsequently, article-level scores are combined across all articles within a given day to construct daily sentiment time series, controlling for publisher-specific variation and calendar effects within a fixed-effects framework.
Since our dataset is similar to that of \textcite{shapiro2022measuring}, albeit covering a different time period, we compare our daily sentiment indices against established benchmarks, including the Shapiro-style measure and lexicon-based indices following the approach of \textcite{barbaglia2024forecasting}. 
As a final step, we validate the alternative sentiment indices through a resource-intensive human-labeling exercise \parencite{ludwig2025large}. Following principles from experimental economics, we recruit a large sample of paid, incentivized annotators via Prolific \parencite{palan2018prolific}, linking compensation to performance.  Annotators are rewarded for assigning ratings within one Likert point of the ex-post consensus label for each item—an output-agreement incentive mechanism designed to discourage low-effort responses when ground-truth labels are costly to obtain \parencite{dasgupta2013crowdsourced}. This validation exercise allows us to assess whether differences across indices reflect only mechanical differences in aggregation or instead correspond to improvements in sentiment classification relative to a human benchmark. 

Empirically, the transformer-based measures differ systematically from the dictionary-based indices. They produce sentiment distributions that are more dispersed and more polarized, with greater mass in both the positive and negative tails. This suggests that context-aware models classify a larger share of text as sentiment-bearing, rather than neutral. The human-labeling exercise confirms this pattern: across several evaluation metrics, transformer-based scores are closer to human annotations than the dictionary-based alternatives. Overall, the results indicate that replicating the benchmark analysis with transformer-based sentiment is useful not only as a robustness exercise, but also as a measurement improvement. The resulting daily mood indices identify sentiment patterns that dictionary-based methods tend to smooth out and therefore differ meaningfully from existing benchmarks.

The remainder of the paper is organized as follows. Section 2 reviews related work on news-based indicators and sentiment extraction. Section 3 describes the data and pre-processing steps. Section 4 introduces the transformer scoring model and the construction of article-level sentiment indices. Section 5 aggregates these measures into daily mood indices and compares them to the Shapiro benchmark. Section 6 presents validation results and additional robustness checks, and Section 7 concludes.

\section{Literature Review on Text-Based Sentiment Indices}

The literature on text-based sentiment indices in macroeconomics follows a broadly common pipeline---data collection, text pre-processing, sentiment/semantic extraction, temporal aggregation, and forecasting integration---but differs substantially in the \emph{extraction} and \emph{aggregation} stages, which are the two main sources of methodological heterogeneity. Most studies build indices from large news archives, with Dow Jones Factiva widely used in Euro Area and Italian applications \parencite{aprigliano2023power, ashwin2024nowcasting, barbaglia2024forecasting, magro2025can}, LexisNexis common in UK and US settings \parencite{rambaccussing2020forecasting, shapiro2022measuring}, and specialized datasets such as Ravenpack and GDELT used in global macro-finance analyses \parencite{gross2024learning, zhang2025interpretable}. Geographic coverage ranges from country-specific indices (e.g., Italy, Spain, Korea, and the Philippines) to Euro Area and cross-country frameworks \parencite{aguilar2021can, arcin2025constructing, seo2024measuring, barbaglia2025sentiment, de2025enhancing, kalamara2022making}. Pre-processing choices are relatively standardized (tokenization, lemmatization, stop-word removal), although multilingual corpora introduce an important trade-off between translation-based harmonization and native-language approaches \parencite{ashwin2024nowcasting, barbaglia2024forecasting, magro2025can}.

The key methodological distinction concerns \emph{how sentiment is extracted}. A first family uses \emph{vocabulary-based} methods, in which texts are scored by matching tokens to predefined lexicons (e.g., positive, negative, or uncertainty terms) and aggregating counts into a sentiment measure. These methods are transparent, computationally light, and easy to scale to very large corpora, which makes them attractive for real-time monitoring and replication \parencite{shapiro2022measuring, kalamara2022making, barbaglia2025sentiment, zhang2018research}. Their main limitation is that they treat words largely in isolation and therefore handle context imperfectly (e.g., negation, intensifiers, semantic ambiguity, and domain-specific meanings), unless dictionaries are carefully adapted; recent work addresses this through valence shifters and macro-specific lexicons such as the Economic Lexicon \parencite{shapiro2022measuring, barbaglia2025sentiment}. A second family uses \emph{transformer-based} models (e.g., FinBERT, language-specific encoders, and LLM-based classifiers), which infer sentiment from contextual embeddings learned through self-attention and typically perform better on complex or ambiguous financial language \parencite{arcin2025constructing, jiang2023financial, ravi2025large, de2025enhancing}. The trade-off is lower interpretability, higher computational cost, and greater dependence on training data, model specification, and fine-tuning choices. A related but more specialized strand combines semantic structure and sentiment through topic--sentiment models; these improve interpretability by linking tone to latent topics, but they add modeling complexity and stronger identification assumptions \parencite{chen2025structural}.

Beyond extraction, a second central challenge is \emph{aggregation}, especially because news is observed at high frequency (often with thousands of articles per day) while macroeconomic targets such as GDP and inflation are monthly or quarterly. Raw article- or sentence-level sentiment is typically too noisy and composition-sensitive to use directly, so studies must choose both a cross-sectional aggregation rule and a temporal smoothing scheme. Common cross-sectional choices include simple averages, length-weighted averages, diffusion ratios, and volume-adjusted scores (e.g., summing sentiment or scaling by log news volume) \parencite{aprigliano2023power, barbaglia2023forecasting, zhang2025interpretable, arcin2025constructing, magro2025can}. Temporal aggregation is handled through rolling windows (e.g., 7- or 30-day averages), month-level fixed effects with outlet controls, or cumulative quarter-to-date measures for GDP nowcasting \parencite{kalamara2022making, seo2024measuring, aguilar2021can, shapiro2022measuring, rambaccussing2020forecasting}. In multi-country settings, sentiment is often first computed at the national level and then aggregated using GDP weights; final indices are typically standardized (z-scores) or rebased to an index level for comparability across time and countries \parencite{ashwin2024nowcasting, magro2025can, aguilar2021can, seo2024measuring}. These aggregation choices are not merely technical: they determine whether the final daily or monthly index captures persistent macroeconomic information or instead reflects short-lived news bursts and changes in source composition.

Finally, sentiment indices are integrated into mixed-frequency forecasting frameworks such as Bayesian Model Averaging, MIDAS, and regularized regressions (e.g., ridge) \parencite{aprigliano2023power, ashwin2024nowcasting, barbaglia2023forecasting, de2025enhancing}. Overall, the literature suggests that vocabulary-based approaches remain strong and interpretable baselines, while transformer-based methods increasingly provide additional predictive gains when contextual interpretation and topic-specific tone are important for the forecasting target \parencite{kalamara2022making}. In the Appendix, a summary is provided in Table \ref{tab:lit_review_readable}.

\section{Data and Construction of Sentiment Indices}


This section describes the construction of the sentiment measures used in the empirical analysis. We first present the news corpus and the sample selection criteria. We then describe the transformer-based approach, which classifies sentiment at the sentence level and aggregates sentence-level predictions into document-level indices. Finally, we introduce vocabulary-based benchmark measures used to compare the transformer-based indicators with a transparent lexicon-based alternative.

\subsection{Dataset: News Corpus}

Our dataset is sourced from Dow Jones Factiva and spans January 2025 to December 2025. We retain English-language news items classified in the finance industry and restrict the sample to documents of type \texttt{article} with non-empty full text. Specifically, we exclude observations with missing or blank article bodies and keep only articles with more than 300 words. The resulting sample contains 143{,}755 articles.

\subsection{Transformer-Based Sentiment Classification and Document-Level Measures}

We measure the sentiment of each article using FinBERT, a BERT-based language model fine-tuned for financial sentiment classification \parencite{araci2019finbert}. FinBERT assigns each text unit to one of three sentiment classes: positive, neutral, or negative. Because BERT-based transformer models such as FinBERT are subject to input-length constraints, typically 512 tokens, direct classification of many full-length financial news articles is not feasible.\footnote{Recent transformer architectures, such as Longformer \parencite{beltagy2020longformer} and BigBird \parencite{zaheer2020big}, can process substantially longer documents through sparse-attention mechanisms. However, their application to financial sentiment analysis would require the development and validation of a dedicated sentiment model, including additional labeled data for fine-tuning. In contrast, FinBERT provides a well-established finance-specific sentiment classifier that can be directly applied at the sentence level. Given the strong performance of transformer-based sentiment models in financial text, we view sentence-level aggregation as a transparent and computationally efficient solution, while leaving the development of long-sequence financial sentiment models for future research.} We therefore adopt a sentence-level approach: each article is segmented into individual sentences, sentiment is classified separately for each sentence, and the resulting sentence-level predictions are subsequently aggregated into document-level sentiment measures. This approach is computationally scalable and allows the construction of alternative sentiment indicators that emphasize different aspects of article tone.

Table~\ref{tab:sentence_stats} reports descriptive statistics for article length and FinBERT sentence-level classifications. On average, articles contain approximately 49 sentences, most of which are classified as neutral, consistent with the largely informational nature of financial news coverage. 
Sentence-level sentiment predictions are aggregated into six alternative transformer-based document-level indicators. Considering multiple aggregation methods allows us to evaluate the robustness of the resulting sentiment dynamics to alternative definitions of article-level tone. The indices emphasize different dimensions of sentiment, including overall tone, polarity among sentiment-bearing sentences, sentiment dominance, nonlinear imbalance, opening-sentence tone, and investor confidence. Detailed definitions, formulas, and illustrative examples are reported in Appendix~\ref{app:sentiment_indices}. The baseline measure, \texttt{index\_stat}, captures net sentiment relative to the total number of sentences, thereby accounting for neutral content and article length. The opening-sentence measure, \texttt{index\_stat\_30}, applies the same logic to the first 30 sentences, capturing the tone conveyed at the beginning of the article. The polarity-based measure, \texttt{index\_pos\_neg}, focuses only on sentiment-bearing sentences by comparing positive and negative classifications while excluding neutral sentences. The dominance measure, \texttt{index\_pos\_neg\_max}, emphasizes the prevailing sentiment direction within the article. The calibrated measure, \texttt{index\_pos\_neg\_tanh}, applies a nonlinear transformation to sentiment imbalance in order to attenuate extreme values. Finally, the Investor Confidence Index, \texttt{index\_ICI}, is based on the log-ratio of positive to negative sentence counts.

\begin{table}[h]
    \centering
    \begin{tabular}{lrrrrrrrr}
    \toprule
   & mean & std & min & 25\% & 50\% & 75\% & max \\
    \midrule
    Number of sentence   & 48.72 & 53.54 & 1.00 & 16.00 & 27.00 & 50.00 & 345.00 \\
    Positive  & 6.36 & 10.87 & 0.00 & 0.00 & 3.00 & 8.00 & 166.00 \\
    Neutral  & 37.23 & 46.15 & 0.00 & 9.00 & 18.00 & 40.00 & 343.00 \\
    Negative & 5.13 & 9.15 & 0.00 & 0.00 & 1.00 & 6.00 & 162.00 \\
    \bottomrule
    \end{tabular}
    
    \caption{Descriptive statistics of sentence counts and sentence-level sentiment labels per article}
    \label{tab:sentence_stats}
\end{table}

\subsection{Vocabulary-Based Benchmark Indices}

As a benchmark, we also construct lexicon-based sentiment indices. In particular, we implement a dictionary-based approach following the aggregation procedure in \textcite{shapiro2022measuring} and use the economic vocabulary developed by \textcite{barbaglia2025sentiment}. Including these measures allows us to compare our transformer-based sentiment indicators with a transparent and widely used alternative in the macro-finance text literature.

The key difference with our transformer-based approach concerns the \emph{unit of aggregation}. In the vocabulary-based method, sentiment is observed at the \emph{word} level only: each token is matched to the lexicon and assigned a polarity (or score), and the document-level indicator is obtained by a simple aggregation of these word-level signals (typically a sum or normalized sum of positive and negative word counts/scores). By contrast, in the transformer-based approach, sentiment is first inferred at the \emph{sentence} level using contextual information, and document sentiment is then constructed by aggregating sentence-level predictions. This distinction is substantive: lexicon-based indices are simpler and highly interpretable, but they cannot directly account for sentence context, whereas transformer-based indices are more flexible but require an additional modeling and aggregation layer.

\begin{figure}[htbp]
    \centering
       \caption{Empirical distributions of document-level sentiment indices. The figure compares the full distributions of transformer-based and vocabulary-based measures, highlighting differences in scale, dispersion, and tail behavior across aggregation methods.}\includegraphics[width=\textwidth,height=1\textheight,keepaspectratio]{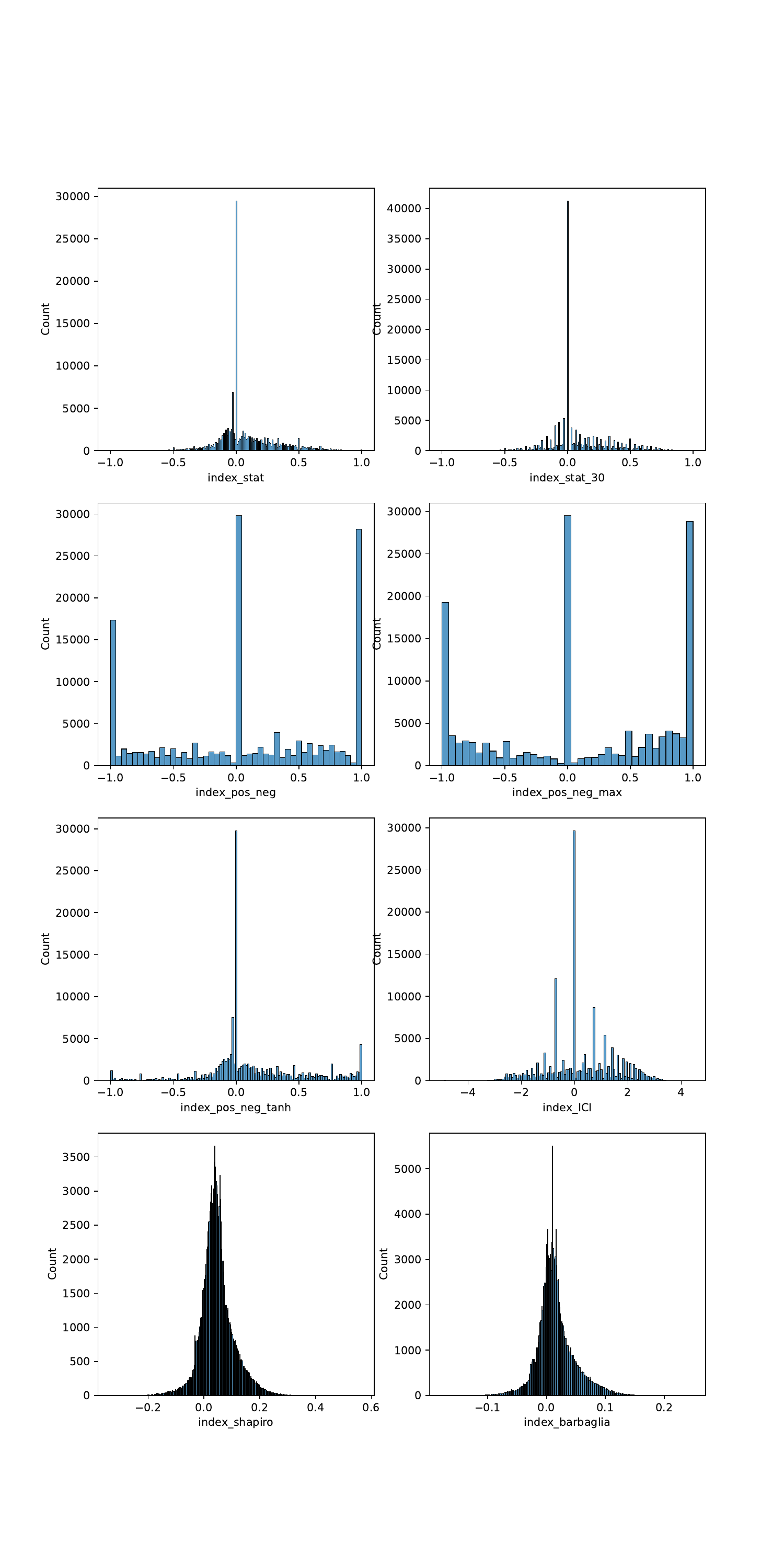}
 
    \label{fig:hist_indices}
\end{figure}

This comparison is important for two reasons. First, lexicon-based indices are computationally lightweight and easy to replicate, making them a natural benchmark for large-scale empirical applications. Second, estimating both approaches on the same corpus allows us to assess whether the richer contextual information embedded in transformer-based sentiment measures translates into materially different signals and, ultimately, improved empirical performance.
Table~\ref{tab:summary_indices} reports descriptive statistics for the main document-level sentiment indices (\(N=143{,}755\)), while Figure~\ref{fig:hist_indices} complements these moments by showing their full empirical distributions. Across all measures, the central tendency is slightly positive (means above zero and medians that are zero or mildly positive), indicating that the corpus is on average weakly tilted toward positive tone. At the same time, both the table and the histograms show substantial heterogeneity in dispersion, driven by differences in aggregation rules and normalization choices.

\begin{table}[]
    \centering  
\begin{tabular}{lrrrrrrrr}
\toprule
 & mean & std & min & 25\% & 50\% & 75\% & max \\
\midrule
index\_stat    & 0.08 & 0.24 & -1.00 & -0.04 & 0.00 & 0.20 & 1.00 \\
index\_stat\_30    & 0.09 & 0.25 & -1.00 & -0.03 & 0.00 & 0.22 & 1.00 \\
index\_pos\_neg     & 0.09 & 0.68 & -1.00 & -0.43 & 0.00 & 0.71 & 1.00 \\
index\_pos\_neg\_max & 0.10 & 0.72 & -1.00 & -0.60 & 0.00 & 0.83 & 1.00 \\
index\_pos\_neg\_tanh  & 0.12 & 0.40 & -1.00 & -0.05 & 0.00 & 0.31 & 1.00 \\
index\_ICI       & 0.22 & 1.27 & -4.98 & -0.69 & 0.00 & 1.10 & 4.44 \\
\midrule
index\_shapiro    & 0.05 & 0.06 & -0.33 & 0.01 & 0.04 & 0.07 & 0.57 \\
index\_barbaglia   & 0.02 & 0.03 & -0.18 & -0.00 & 0.01 & 0.03 & 0.25 \\
\bottomrule
\end{tabular}
\caption{Descriptive statistics of document-level sentiment indices (\(N=143{,}755\)). The table reports summary moments for transformer-based and vocabulary-based measures, highlighting differences in central tendency, dispersion, and support across aggregation methods.}    \label{tab:summary_indices}
\end{table}

\paragraph{Vocabulary-Based versus Transformer-Based Sentiment Measures}

The distributional evidence in Table~\ref{tab:summary_indices} and
Figure~\ref{fig:hist_indices}  highlights a clear contrast between the benchmark vocabulary-based indices and the transformer-based measures. The benchmark vocabulary-based indices (\texttt{index\_shapiro} and \texttt{index\_barbaglia}) are concentrated in a very narrow interval around zero, with low standard deviations (\(0.06\) and \(0.03\), respectively), indicating smooth but low-amplitude variation under word-level aggregation. The polarity-conditional indices (\texttt{index\_pos\_neg} and \texttt{index\_pos\_neg\_max}), by contrast, are much more dispersed (\(\text{std}=0.68\) and \(0.72\)) and place substantial mass toward the bounds in the histograms, reflecting stronger sensitivity to polarized content because neutral sentences are excluded. The sentence-normalized indices (\texttt{index\_stat} and \texttt{index\_stat\_30}) are more compressed (\(\text{std}\approx 0.24\)--\(0.25\)) and centered near zero, consistent with neutral-content dilution. The tanh-calibrated index (\texttt{index\_pos\_neg\_tanh}) lies between these cases (\(\text{std}=0.40\)), showing that nonlinear scaling attenuates extreme values while preserving directional variation. Finally, \texttt{index\_ICI} has the widest support and the largest dispersion (\(\text{std}=1.27\), range \([-4.98,\,4.44]\)), as expected from its unbounded log-ratio construction. Overall, the distributional evidence confirms that the choice of aggregation method materially affects the scale, tail behavior, and interpretability of the resulting sentiment signal.
\clearpage
\thispagestyle{empty}

\section{Daily Mood Indices: Construction and Comparison}
\label{sec:estimation_daily_mood}

This section describes how the article-level sentiment measures introduced above are aggregated into daily mood indices. The goal is to construct smooth and interpretable time series that capture systematic variation in sentiment over the calendar year while accounting for heterogeneity in publication volume and persistent differences across publishers. Although the underlying sentiment measures differ in their treatment of neutrality, dominance, and nonlinear scaling, they are all aggregated using the same procedure, which makes the resulting daily indices directly comparable.

For each sentiment index \(k\), let \(y_{ijd}^{(k)}\) denote the article-level sentiment value for article \(j\), published by outlet \(i\), on calendar day-of-year \(d\). To recover the average sentiment associated with each day net of publisher-specific sentiment levels, we estimate the following fixed-effects regression:
\begin{equation}
y_{ijd}^{(k)} \;=\; \alpha_i^{(k)} \;+\; \sum_{d' \in \mathcal{D}} \beta_{d'}^{(k)} \mathbf{1}\{d=d'\} \;+\; \varepsilon_{ijd}^{(k)},
\end{equation}
where \(\alpha_i^{(k)}\) captures time-invariant outlet-specific differences in sentiment and \(\beta_{d'}^{(k)}\) is the average sentiment associated with calendar day \(d'\). Estimation uses analytic weights proportional to publication volume so that days with more content receive greater influence. Standard errors are clustered at the publisher level to allow for arbitrary within-outlet dependence over time.

The estimated day-of-year coefficients \(\{\widehat{\beta}_d^{(k)}\}_{d\in\mathcal{D}}\) define the raw daily mood series for index \(k\). Because day fixed effects are identified only up to an additive constant, we recenter each series to have mean zero across observed days. Let
\begin{equation}
\bar{\beta}^{(k)} = \frac{1}{|\mathcal{D}|}\sum_{d\in\mathcal{D}}\widehat{\beta}_d^{(k)},
\end{equation}
and define the normalized daily mood index as
\begin{equation}
\mathrm{Mood}_d^{(k)} = \widehat{\beta}_d^{(k)} - \bar{\beta}^{(k)}.
\end{equation}
For each day \(d\), we also compute a \(95\%\) confidence interval,
\[
\widehat{\beta}_d^{(k)} \pm 1.96\,\mathrm{SE}\!\left(\widehat{\beta}_d^{(k)}\right),
\]
and apply the same recentering transformation to the lower and upper bounds. This normalization ensures that the daily mood profile captures deviations from the annual average sentiment rather than differences relative to an arbitrary omitted day.

\begin{figure}[h!]
    \centering
    \begin{subfigure}[b]{0.9\linewidth}
        \centering
        \includegraphics[width=\linewidth]{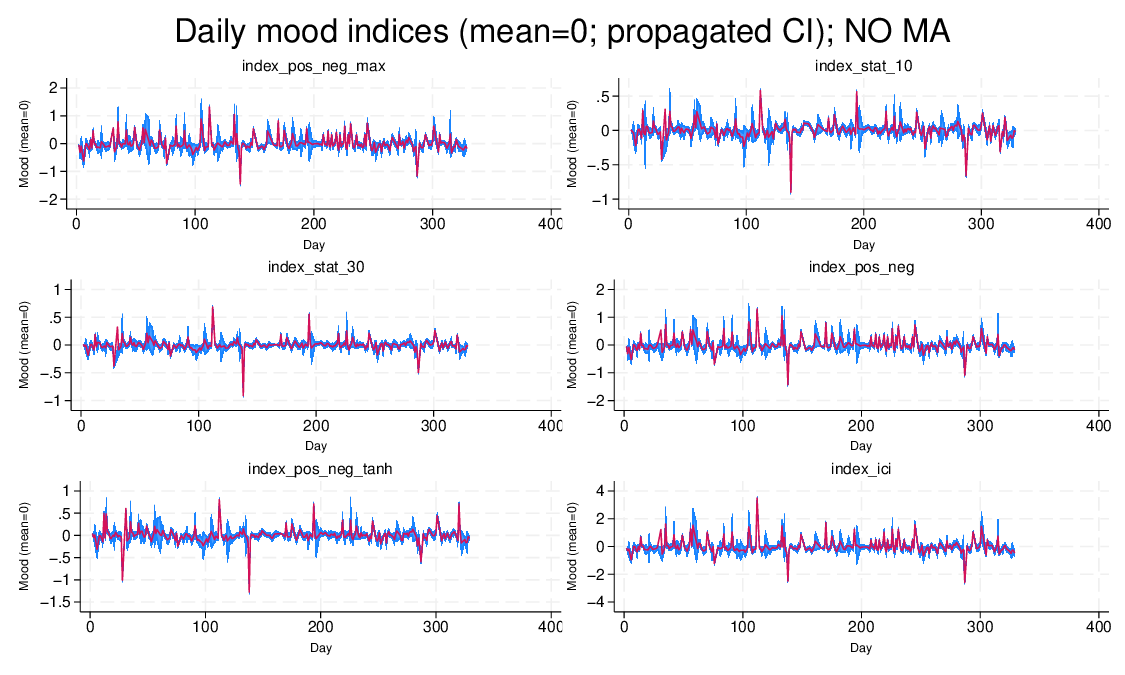}
        \caption{Daily mood indices (raw)}
        \label{fig:transformer}
    \end{subfigure}

    \begin{subfigure}[b]{0.9\linewidth}
        \centering
        \includegraphics[width=\linewidth]{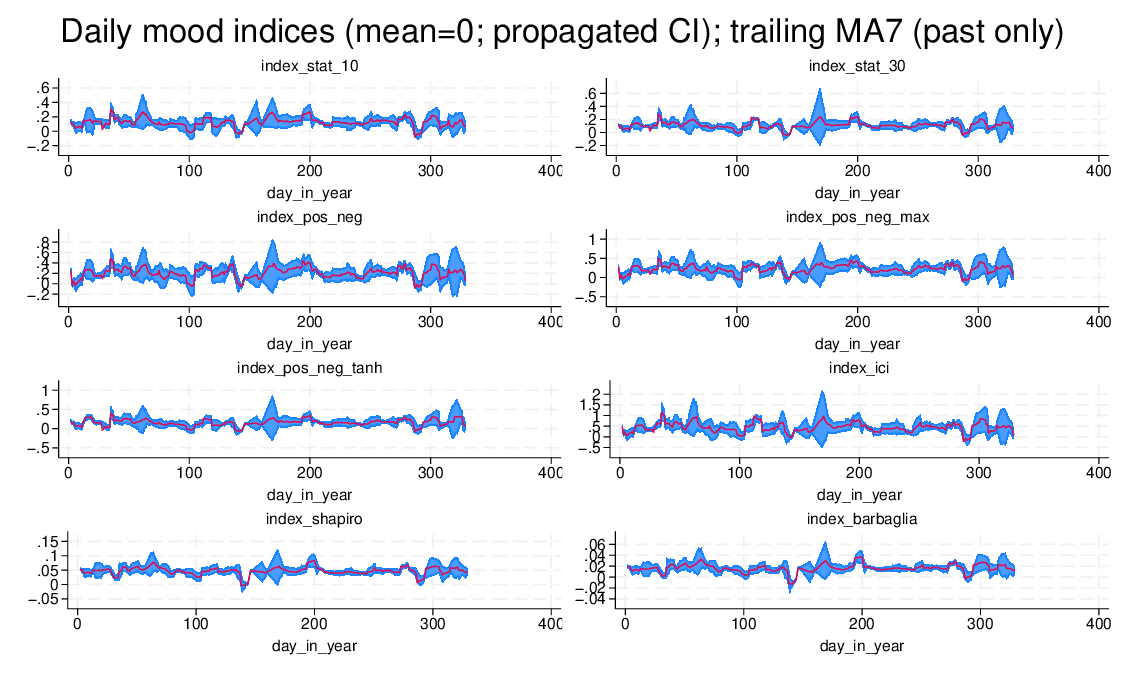}
        \caption{Daily mood indices (7-day moving average)}
        \label{fig:transformer_ma}
    \end{subfigure}

    \caption{Comparison of alternative daily mood indices}
    \label{fig:index_comparison}
\end{figure}

For visualization and cross-index comparison, we report both the raw daily mood series and a seven-day moving-average version (with corresponding confidence bands), as shown in Figure~\ref{fig:index_comparison}. Panel~\ref{fig:transformer} displays the raw transformer-based indices, which exhibit strong co-movement and similar timing of peaks and troughs despite differences in scale and normalization, while Panel~\ref{fig:transformer_ma} adds the vocabulary-based indices and applies seven-day smoothing to all series, reducing short-run volatility and making common dynamics more visible. The figure also highlights the much narrower support of the vocabulary-based indices relative to the polarity-based transformer measures. These daily mood profiles are obtained from the nonparametric day-of-year fixed-effects specification, which allows sentiment to vary flexibly over the calendar year without imposing a parametric dynamic structure. For each day \(d\), we compute coverage diagnostics (number of observations and total analytic weight, i.e., publication volume) and flag days below the 5th percentile of the annual weight distribution as low-coverage; these days are retained but explicitly identified in graphical and tabular outputs. Indices are reported only for days with sufficient underlying observations to enter the estimation sample (days with no usable data are omitted), and all series are recentered to have mean zero so that they are directly comparable across specifications. Overall, despite differences in normalization and smoothness, the indices display strong co-movement, indicating that the main daily sentiment dynamics are robust to the choice of document-level aggregation method.

\subsection{Comovement Across Daily Mood Indices}
\label{sec:correlations}

To assess the internal consistency of the daily mood measures and quantify the extent to which alternative constructions capture common sentiment dynamics, we compute pairwise correlations across the six daily indices. We report contemporaneous correlations (lag~0) as well as lead--lag correlations at one- and two-day horizons, defined as $\mathrm{corr}(X_t,Y_t)$, $\mathrm{corr}(X_t,Y_{t-1})$, and $\mathrm{corr}(X_t,Y_{t-2})$. The lead--lag correlations help detect whether differences in normalization or calibration induce systematic timing shifts in measured sentiment. Correlations are computed pairwise using all available overlapping observations for each index pair and lag, and we store both the correlation coefficient and the corresponding sample size.

\begin{table}[!htbp]
\centering
\footnotesize            
\setlength{\tabcolsep}{3pt}
\renewcommand{\arraystretch}{0.9}
\caption{Summary of pairwise correlations $\rho$ by lag (among the 6 indices, excluding Shapiro/Barbaglia)}
\label{tab:rho_summary_by_lag_core6_only}

\begin{tabular}{rrrrrrrrrrrr}
\hline
Lag & Pairs & Mean & SD & Median & Min & Max & Mean$|\rho|$ & Med$|\rho|$ &
$\%(|\rho|\ge0.7)$ & $\%(|\rho|\ge0.8)$ & $\%(|\rho|\ge0.9)$\\
\hline
0 & 15 & 0.903 & 0.045 & 0.895 & 0.835 & 0.993 & 0.903 & 0.895 & 100.0 & 100.0 & 40.0\\
1 & 15 & 0.741 & 0.035 & 0.746 & 0.685 & 0.797 & 0.741 & 0.746 & 93.3 & 0.0 & 0.0\\
2 & 15 & 0.645 & 0.038 & 0.630 & 0.592 & 0.714 & 0.645 & 0.630 & 6.7 & 0.0 & 0.0\\
\hline
\end{tabular}
\end{table}

Table~\ref{tab:rho_summary_by_lag_core6_only} summarizes the distribution of pairwise correlations among the mood indices at different temporal lags. At lag~0, contemporaneous correlations are very high across all index pairs, with a mean correlation of \(0.903\) , and all pairs exhibiting  \(|\rho|\ge 0.8\); \(40\%\) of pairs also exceed \(|\rho|\ge 0.9\). This indicates strong same-day comovement across indices constructed from the same underlying information set. Correlations remain positive but decline at lag~1 (mean \(0.741\), with \(93.3\%\) of pairs still above \(|\rho|\ge 0.7\). By lag~2, correlations decrease further (mean \(0.645\), and only \(6.7\%\) of pairs remain above \(|\rho|\ge 0.7\). Overall, the pattern points to strong contemporaneous comovement and a rapid decay in cross-index correlation as the temporal lag increases.

As an additional consistency check, we compare the two vocabulary-based daily mood indices---\texttt{index\_shapiro} and \texttt{index\_barbaglia}---with the six transformer-based indices. Table~\ref{tab:rho_summary_by_lag_barbaglia_vs_core6} reports summary statistics of the pairwise correlations between each vocabulary-based index and the six transformer-based measures at lags \(0\), \(1\), and \(2\) (where lag 1 corresponds to \(\mathrm{corr}(L_t, X_{t-1})\)).

The results show contemporaneous alignment between vocabulary-based and transformer-based measures, but weaker than the correlations observed within the transformer family. At lag~0, both vocabulary-based indices are moderately to strongly correlated with the six transformer indices: for \texttt{index\_barbaglia}, the mean correlation is \(0.765\) (range \([0.706,\,0.840]\)); for \texttt{index\_shapiro}, the mean correlation is \(0.769\) (range \([0.719,\,0.826]\)). In both cases, all six pairs exceed \(|\rho|\ge 0.7\), but only half exceed \(|\rho|\ge 0.8\), and none exceed \(|\rho|\ge 0.9\). This contrasts with the much tighter contemporaneous comovement among transformer-based indices, where same-day correlations are substantially higher on average and often close to the upper bound. Correlations remain positive at lag~1 but decay noticeably. For \texttt{index\_barbaglia}, the mean lagged correlation falls to \(0.658\) (with only \(16.7\%\) of pairs above \(|\rho|\ge 0.7\)); for \texttt{index\_shapiro}, it is slightly higher at \(0.670\) (with \(50.0\%\) of pairs above \(|\rho|\ge 0.7\)). By lag~2, correlations decline further to \(0.616\) for \texttt{index\_barbaglia} and \(0.631\) for \texttt{index\_shapiro}, and no pair remains above \(|\rho|\ge 0.7\).

Overall, the evidence suggests that the vocabulary-based indices capture the same broad daily sentiment dynamics as the transformer-based measures (especially contemporaneously), but with systematically lower alignment and a more compressed signal. This is consistent with the earlier distributional evidence: vocabulary-based indices are smoother and lower-amplitude because they aggregate word-level polarity, whereas transformer-based indices extract sentiment at the sentence level and are more sensitive to contextual polarity and sentiment intensity.

\begin{table}[!htbp]
\centering
\footnotesize 
\setlength{\tabcolsep}{3pt}      
\renewcommand{\arraystretch}{0.9}
\caption{Summary of pairwise correlations $\rho$ by lag}
\label{tab:rho_summary_by_lag_barbaglia_vs_core6}

\begin{tabular}{rrrrrrrrrrrr}
\hline
Lag & Pairs & Mean & SD & Median & Min & Max & Mean$|\rho|$ & Med$|\rho|$ &
$\%(|\rho|\ge0.7)$ & $\%(|\rho|\ge0.8)$ & $\%(|\rho|\ge0.9)$\\
\hline
\multicolumn{12}{c}{Barbaglia vs each of the 6 indices}\\ 
\hline
0 & 6 & 0.765 & 0.062 & 0.761 & 0.706 & 0.840 & 0.765 & 0.761 & 100.0 & 50.0 & 0.0\\
1 & 6 & 0.658 & 0.054 & 0.652 & 0.608 & 0.725 & 0.658 & 0.652 & 16.7 & 0.0 & 0.0\\
2 & 6 & 0.616 & 0.035 & 0.613 & 0.580 & 0.661 & 0.616 & 0.613 & 0.0 & 0.0 & 0.0\\
\hline
\multicolumn{12}{c}{Shapiro vs each of the 6 indices}\\
\hline
0 & 6 & 0.769 & 0.051 & 0.769 & 0.719 & 0.826 & 0.769 & 0.769 & 100.0 & 50.0 & 0.0\\
1 & 6 & 0.670 & 0.044 & 0.671 & 0.623 & 0.723 & 0.670 & 0.671 & 50.0 & 0.0 & 0.0\\
2 & 6 & 0.631 & 0.024 & 0.635 & 0.597 & 0.659 & 0.631 & 0.635 & 0.0 & 0.0 & 0.0\\
\hline
\end{tabular}
\end{table}

As a final consistency check, we examine whether the daily mood indices display
mechanical persistence or systematic calendar patterns that could affect the
interpretation of the moving-average series. The results, reported in Appendix~\ref{app:serial_weekday}, show only modest short-run autocorrelation and little persistence beyond the first lag. Residualizing the daily series with respect to weekday fixed effects leaves the main pattern essentially unchanged, suggesting that the smoothed indices are not primarily driven by serial dependence in daily sentiment. At the same time, daily sentiment displays some heterogeneity across weekdays, which motivates the use
of calendar controls when constructing and interpreting the daily mood measures.

\section{Human Labeling and Validation Sample}
\label{sec:labeling}

The preceding sections show that transformer-based and vocabulary-based measures produce related but distinct sentiment signals. However, differences in dispersion, persistence, or comovement do not by themselves establish which measure better captures the tone perceived by human readers. This is particularly important in financial news, where sentiment often depends on context, negation, and the economic meaning of a statement rather than on isolated positive or negative words. We therefore validate the alternative text-based sentiment measures against independent human judgments.

To do so, we conducted a large-scale annotation exercise on a sample of 588 newspaper articles drawn from the corpus \footnote{The annotation experiment was preregistered prior to data collection on the Open Science Framework (OSF). The preregistration, including the experimental design, recruitment procedure, and analysis plan, is available at \url{https://osf.io/5ruva/overview}.}. Participants were recruited through \textit{Prolific} \parencite{palan2018prolific} and each evaluated four randomly assigned articles. Each article was independently rated by exactly three annotators, yielding 1,776 sentiment evaluations from 444 unique participants.\footnote{A pilot study was conducted to test instructions, timing, and survey functionality. Pilot observations were excluded from all analyses. Eligibility for the main study was restricted to participants located in the United Kingdom or the United States who reported English as their first language. Individuals who participated in the pilot were excluded from the main study.}

Participants assessed the sentiment expressed in each article on a five-point Likert scale ranging from 1 (\textit{Very Negative}) to 5 (\textit{Very Positive}). The instructions emphasized that sentiment refers to the tone and emotional content of the language, rather than the economic substance of the article. In addition, participants were asked to predict the sentiment rating that most other participants would assign to the same article. This second task elicited beliefs about social consensus and was incentivized using an output-agreement mechanism \parencite{dasgupta2013crowdsourced}. Participants received a base payment of £1.40 and could earn a bonus of £0.10 per article if their prediction was within one Likert point of the realized consensus rating. The average bonus payment was £0.195 (SD = 0.113), ranging from £0 to £0.40.

To promote independent evaluations, copy-and-paste functionality was disabled and participants were instructed not to leave or refresh the survey page during the task.\footnote{Leaving the survey page resulted in the loss of eligibility for the performance-based bonus. These restrictions were implemented to reduce the use of external tools and ensure comparable annotation conditions across participants.} After data collection, ratings were aggregated at the article level to construct consensus sentiment labels. These labels serve as the benchmark against which automated sentiment measures are evaluated.

\begin{figure}
    \centering
    \includegraphics[width=1\linewidth]{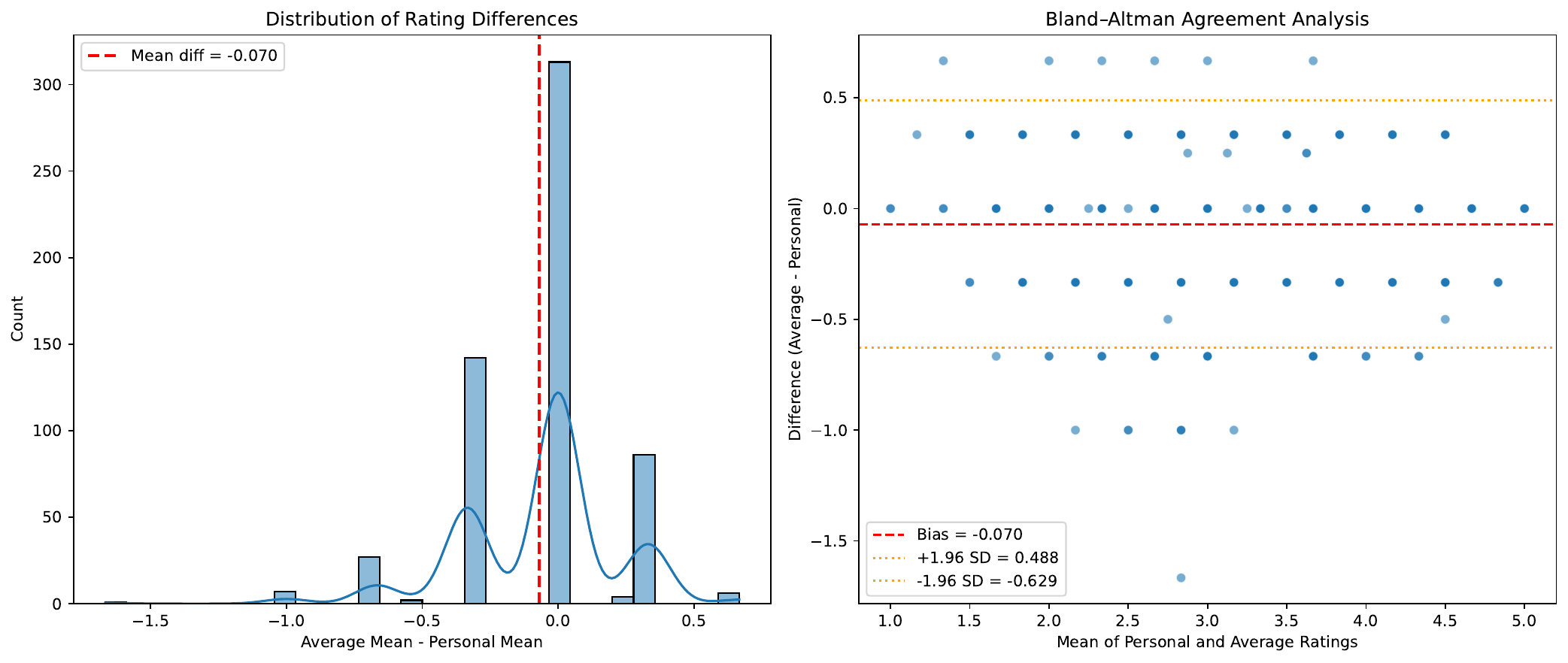}
    \caption{Inter-rater agreement and rating consistency. The figure shows the distribution of individual sentiment ratings, their deviation from article-level averages, and the overall agreement structure across the three independent evaluators per article.}
    \label{fig:quality}
\end{figure}

To assess data quality and the reliability of the human sentiment measure, we examine the agreement between raters. The distribution of ratings shows strong consistency between individual evaluations and article-level averages. The maximum deviation between any individual rating and the mean of the corresponding article is less than one point on the scale of 1--5. The average difference between individual ratings and article means is -0.07, indicating that there is no systematic bias in individual judgments relative to the aggregated measure. A Bland-Altman style analysis \footnote{A Bland--Altman analysis \parencite{bland1999measuring} examines agreement by plotting the differences between paired measurements against their mean, allowing the identification of systematic bias and dispersion.} further confirms high agreement, with most observations falling within a $\pm 0.5$ point range between individual and aggregated ratings. Given that each article is independently rated exactly three times, these results indicate that the aggregation is based on a sufficiently dense and balanced structure to ensure a stable measurement of sentiment at the article level. Overall, the high level of inter-rater agreement supports the use of the averaged rating across the three evaluators as a reliable ground-truth sentiment measure.The plots are reported in Figure \ref{fig:quality}, which summarizes the agreement between individual ratings and aggregated sentiment at the article-level, as well as the distribution of the rating differences between the evaluators.

\section{Validation Against Human Judgments}

The human-labeling exercise described in Section~\ref{sec:labeling} establishes
an external benchmark, or ground truth, against which alternative sentiment
measures can be now  evaluated. This benchmark was necessary because the previous
sections show that transformer-based and vocabulary-based indices recover related
but distinct signals, while differences in scale, dispersion, or time-series behavior
do not by themselves reveal which measure better captures the tone perceived by
readers. This section therefore uses three complementary exercises the consensus human labels to assess whether
transformer-based measures more closely approximate human judgments of
financial-news sentiment than vocabulary-based benchmarks. This exercise parallels the correlation analysis in Section~\ref{sec:correlations}, but focuses on the labeled validation subsample and replaces cross-index comovement with agreement relative to the human
benchmark. It therefore provides a simple measure of external validity and allows
us to compare the degree to which alternative indices track human perceptions of
sentiment.
Second, we use each sentiment index separately to predict the human sentiment
labels in-sample, both in regressions and in classification models. This allows us
to assess whether each automated measure explains human ratings and correctly
classifies articles as negative, neutral, or positive.
Third, we evaluate out-of-sample performance using train-test splits. We estimate
regression and classification models on a training sample and test them on held-out
observations. This provides a stricter validation exercise by showing whether each
sentiment measure generalizes to unseen labeled texts rather than only fitting the
labeled sample.

Together, these exercises assess how closely the automated measures reproduce
human sentiment judgments and whether transformer-based methods improve on
vocabulary-based benchmarks.

\subsection{Correlation with Human Sentiment Judgments}

\begin{figure}
    \centering
    \includegraphics[width=0.9\linewidth]{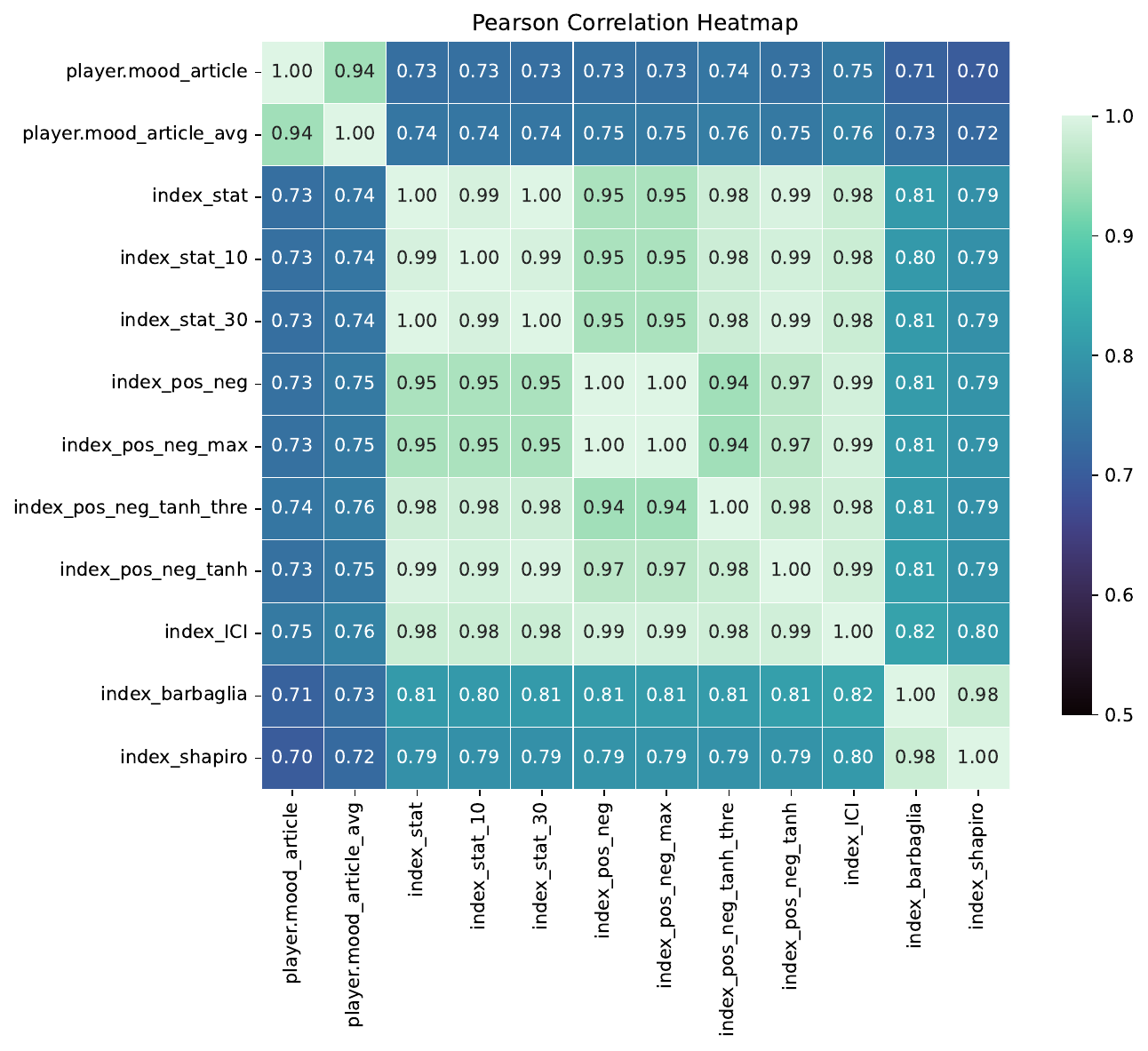}
    \caption{Pearson Correlation Heatmap }
    \label{fig:corrpearson}
\end{figure}

Figure~\ref{fig:corrpearson} presents the correlation matrix for human sentiment
ratings and the alternative sentiment indices. Consistent with the strong
comovement documented in Section~\ref{sec:correlations}, the automated measures
remain highly correlated with one another in the labeled validation sample. The
purpose of this exercise, however, is different: rather than assessing internal
consistency among indices, it uses the consensus human labels as an external
benchmark for evaluating measurement quality.

All automated measures are positively and substantially correlated with the human
benchmark, but transformer-based indices consistently display the strongest
correlations. This is consistent with the view that contextual language models
better capture semantic nuance in financial news.\footnote{The highest correlations
with the consensus human rating are obtained by the transformer-based indices
(approximately $r=0.75$--$0.76$), compared with $r=0.71$--$0.70$ for the
vocabulary-based measures.}

The matrix also exhibits a pronounced block structure (see green block in Figure~\ref{fig:corrpearson} ). Human ratings form a
distinct cluster ($r=0.94$), transformer-based measures display very high internal
consistency ($r=0.95$--$0.99$), and vocabulary-based indices are themselves highly
correlated ($r=0.98$). These clusters suggest that different methodologies recover
a common sentiment signal while relying on distinct measurement approaches. At the same time, the transformer-based measures display slightly higher correlations with the human benchmark (approximately 0.75 versus 0.70 for the vocabulary-based measures). This first descriptive comparison suggests a stronger association between transformer-based sentiment scores and human judgments. However, correlation alone does not establish the relative quality of the measures, motivating the regression and classification analyses that follow.

\subsection{In-sample analysis}

In this section, we assess in-sample agreement with human sentiment labels using three complementary exercises: multiclass classification performance, confusion matrices, and univariate regressions of human ratings on each automated sentiment measure.

Table \ref{tab:classification_insample} reports the in-sample classification performance of the alternative sentiment measures on the full sample of 588 articles.\footnote{For comparability across sentiment measures, continuous sentiment scores were converted into three sentiment categories (negative, neutral, and positive) using symmetric thresholds of $\pm 0.33$. Scores below $-0.33$ were classified as negative, scores above $0.33$ as positive, and the remaining observations as neutral.} The results reveal a clear distinction between transformer-based and vocabulary-based approaches. All transformer-based indices achieve similar classification performance, with accuracies ranging from 66\% to 68\%, balanced accuracies between 0.67 and 0.69, and macro-F1 scores between 0.67 and 0.69. We focus primarily on the macro-F1 score, as it is the harmonic mean of precision and recall computed separately for each class and then averaged across classes. Unlike overall accuracy, this metric accounts for both false positives and false negatives while assigning equal weight to each sentiment category, making it particularly appropriate for evaluating multiclass sentiment classification.

By contrast, the vocabulary-based indices obtain a macro-F1 score of only 0.187, indicating that they perform poorly across the three sentiment categories. Overall, the results suggest that incorporating contextual information through transformer-based language models substantially improves the ability to reproduce the sentiment judgments of human annotators compared with traditional vocabulary-based methods.

\begin{table}[h]
\centering
\caption{In-sample classification performance against human sentiment labels}
\label{tab:classification_insample}
\begin{tabular}{lccc}
\toprule
Sentiment Measure & Accuracy & Balanced Accuracy & Macro-F1 \\
\midrule
\multicolumn{4}{l}{\textit{Transformer-based measures}} \\
\quad index\_stat           & 0.662 & 0.670 & 0.668 \\
\quad index\_stat\_10       & 0.668 & 0.678 & 0.675 \\
\quad index\_stat\_30       & 0.662 & 0.670 & 0.668 \\
\quad index\_pos\_neg       & \textbf{0.682} & \textbf{0.693} & \textbf{0.687} \\
\quad index\_pos\_neg\_max  & \textbf{0.682} & \textbf{0.693} & \textbf{0.687} \\
\quad index\_pos\_neg\_tanh & \textbf{0.682} & \textbf{0.693} & \textbf{0.687} \\
\quad index\_ICI            & \textbf{0.682} & \textbf{0.693} & \textbf{0.687} \\
\midrule
\multicolumn{4}{l}{\textit{Vocabulary-based measures}} \\
\quad index\_barbaglia      & 0.390 & 0.333 & 0.187 \\
\quad index\_shapiro        & 0.390 & 0.333 & 0.187 \\
\bottomrule
\end{tabular}

\vspace{0.2cm}
\begin{minipage}{0.9\linewidth}
\footnotesize
\textit{Notes:} Human sentiment labels are constructed from the consensus rating of annotators. Continuous sentiment scores are converted into three classes (negative, neutral, positive) using symmetric thresholds of $\pm0.33$. Accuracy denotes the share of correctly classified articles, balanced accuracy averages recall across classes, and Macro-F1 is the unweighted average of class-specific F1 scores.
\end{minipage}
\end{table}

\begin{figure}
    \centering
    \includegraphics[width=0.9\linewidth]{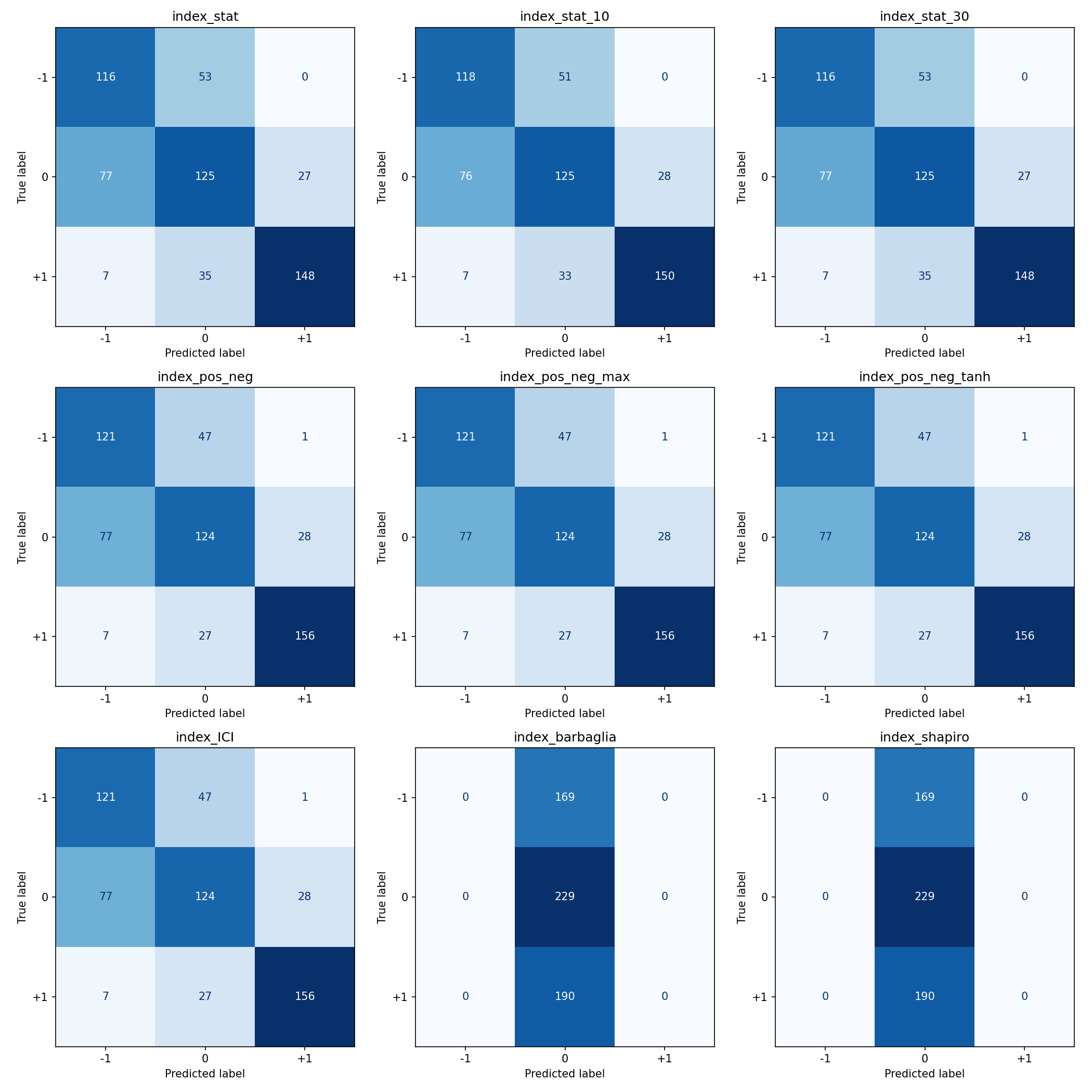}
    \caption{Confusion matrices by sentiment measure.
Rows correspond to human sentiment labels and columns to predicted labels. Transformer-based measures show substantial agreement with human evaluations across all sentiment categories, whereas vocabulary-based measures classify most articles as neutral and exhibit limited ability to distinguish positive from negative sentiment.}
    \label{fig:conf}
\end{figure}

Figure \ref{fig:conf} reports the confusion matrices for the alternative sentiment measures. A clear contrast emerges between transformer-based and vocabulary-based approaches. Transformer-based indices successfully recover all three sentiment categories and exhibit a strong concentration of observations along the main diagonal, indicating substantial agreement with human labels. Their classification errors are primarily local: negative and positive articles are occasionally classified as neutral, while direct confusion between negative and positive articles is rare. This pattern suggests that transformer-based measures capture the direction of sentiment well, with most disagreements arising for articles whose sentiment is relatively weak or ambiguous. In comparison, the vocabulary-based indices display a degenerate classification pattern. Both the Barbaglia and Shapiro measures assign virtually all articles to the neutral category, contrary to the human labelling. As a result, they correctly identify a large share of neutral articles but fail entirely to distinguish between positive and negative sentiment, producing no correct classifications in either extreme category. The confusion matrices therefore illustrate that the superior performance of transformer-based measures stems not only from higher overall accuracy but also from their ability to discriminate across the full sentiment spectrum, whereas vocabulary-based approaches largely collapse sentiment variation into a single neutral class.

Finally, Table \ref{tab:regression_insample} reports univariate OLS regressions of the consensus human sentiment rating on each automated sentiment measure. All coefficients are positive and highly statistically significant ($p<0.001$), indicating that higher automated sentiment scores are consistently associated with more positive human evaluations. The explanatory power of the models is substantial, with $R^2$ values ranging from 0.52 to 0.58. Among the transformer-based measures, the ICI index delivers the strongest performance ($R^2 = 0.581$)
Vocabulary-based indices exhibit slightly lower explanatory power with $R^2$ values of 0.534 for the Barbaglia index and 0.518 for the Shapiro index. Overall, the regression results confirm the findings from the correlation and classification analyses: while all measures capture meaningful variation in human sentiment judgments, transformer-based approaches provide the closest approximation to the human benchmark.

\begin{table}[]
\centering
\caption{In-sample regression performance against human sentiment labels}
\label{tab:regression_insample}
\begin{tabular}{lccc}
\toprule
Sentiment Measure & $R^2$ & Adj. $R^2$ & AIC \\
\midrule
\multicolumn{4}{l}{\textit{Transformer-based measures}} \\
\quad index\_stat           & 0.551 & 0.550 & 195.27 \\
\quad index\_stat\_10       & 0.553 & 0.552 & 192.76 \\
\quad index\_stat\_30       & 0.551 & 0.550 & 195.27 \\
\quad index\_pos\_neg       & 0.560 & 0.560 & 182.99 \\
\quad index\_pos\_neg\_max  & 0.560 & 0.560 & 182.84 \\
\quad index\_pos\_neg\_tanh & 0.558 & 0.557 & 186.16 \\
\quad index\_ICI            & \textbf{0.581} & \textbf{0.581} & \textbf{154.28} \\
\midrule
\multicolumn{4}{l}{\textit{Vocabulary-based measures}} \\
\quad index\_barbaglia      & 0.534 & 0.533 & 217.46 \\
\quad index\_shapiro        & 0.518 & 0.517 & 236.87 \\
\bottomrule
\end{tabular}

\vspace{0.2cm}
\begin{minipage}{0.9\linewidth}
\footnotesize
\textit{Notes:} The table reports univariate OLS regressions of the consensus human sentiment rating on each automated sentiment measure. Higher $R^2$ values indicate stronger agreement with human evaluations, while lower AIC values indicate better model fit. All estimated coefficients are positive and statistically significant at the 1\% level.
\end{minipage}
\end{table}

\subsection{Out-of-Sample Validation}

To assess the external validity of the sentiment measures, in this section we conduct an out-of-sample prediction exercise using a temporal train--test split. Articles are ordered chronologically and divided into training and test samples. The training sample is used for model selection and hyperparameter tuning, while all performance metrics are reported on a hold-out test set containing unseen observations. Model selection is performed using three-fold time-series cross-validation (\texttt{TimeSeriesSplit}) \footnote{We use a time-series split (TimeSeriesSplit from scikit-learn) instead of standard k-fold cross-validation because our data has a natural temporal structure. Newspaper sentiment and human mood ratings are potentially time-dependent, meaning that observations closer in time are more likely to be correlated. Standard k-fold cross-validation would randomly shuffle observations and allow information from the future to leak into the training set, leading to overly optimistic and unrealistic performance estimates.

To preserve the temporal ordering of the data, we implement a three-fold rolling time-series split. In this procedure, the data are ordered chronologically, and each fold uses earlier observations for training and later observations for testing. Specifically, in fold 1 the model is trained on the earliest segment of the data and tested on the next time block; in fold 2 the training window is expanded to include both earlier segments and tested on a later block; and in fold 3 the model is trained on all previous data except the final segment, which is used as the test set.

This structure ensures that the model is always evaluated on genuinely unseen future observations relative to the training data, thereby providing a more realistic assessment of out-of-sample predictive performance. It also allows us to evaluate the stability of transformer-based and dictionary-based sentiment measures across different time periods, reducing the risk that results are driven by a single temporal split.}, ensuring that validation observations always occur after the corresponding training observations. This design prevents look-ahead bias and mimics a realistic forecasting environment. Hyperparameters are selected using grid search with macro-F1 as the optimization criterion.

We consider three classification algorithms. First, a Balanced Random Forest classifier is estimated with the number of trees $\in {100,200}$), maximum tree depth  $\in {3,4,5}$), minimum split size $\in {2,5}$), and minimum leaf size  $\in {1,2}$) selected via cross-validation. Second, we estimate a Ridge classifier with regularization parameter $\alpha \in {0.1,1,10,100}$. Third, we estimate a Support Vector Machine (SVM) classifier with penalty parameter $C \in {0.01,0.1,1,10}$, kernel type (\texttt{linear}, \texttt{rbf}), and kernel parameter $\gamma \in {\texttt{scale}, \texttt{auto}, 0.1, 1}$. All models are trained on the same three-class sentiment labels used in the in-sample classification exercise.

\paragraph{Out-of-Sample Classification Performance}

Table \ref{tab:oos_classification} reports classification performance on the hold-out test sample. Among the machine-learning models, the SVM achieves the highest macro-F1 score (0.627), while the Ridge classifier attains the highest accuracy (0.627) and balanced accuracy (0.639). The Balanced Random Forest exhibits the strongest in-sample performance but experiences a larger decline on the test set, suggesting some degree of overfitting. Across all specifications, positive articles remain the easiest category to classify, whereas neutral articles continue to be the most difficult.

A notable result is that the best-performing transformer-based sentiment indices perform at least as well as the more complex machine-learning classifiers. The \texttt{index\_pos\_neg}, and \texttt{index\_ICI} measures achieve a macro-F1 score of 0.638 and an accuracy of 0.627, matching or slightly exceeding the predictive performance of the Ridge and SVM classifiers. This finding suggests that the transformer-derived sentiment measures already capture most of the information relevant for reproducing human sentiment judgments. Additional machine-learning layers provide only limited incremental gains.

In contrast, the vocabulary-based measures perform poorly out of sample. Both the Barbaglia and Shapiro indices achieve a balanced accuracy of 0.33 and a macro-F1 score of 0.16, indicating performance close to random classification. Taken together, the results confirm the superiority of transformer-based approaches and demonstrate that their advantage persists when evaluated on unseen articles.

\begin{table}[htbp]
\centering
\caption{Out-of-sample classification performance}
\label{tab:oos_classification}
\begin{tabular}{lccc}
\toprule
Method & Accuracy & Balanced Accuracy & Macro-F1 \\
\midrule
\multicolumn{4}{l}{\textit{Machine-learning models}} \\
Balanced Random Forest & 0.593 & 0.601 & 0.611 \\
Ridge Classifier       & 0.627 & \textbf{0.639} & 0.623 \\
SVM                    & 0.610 & 0.619 & 0.627 \\
\midrule
\multicolumn{4}{l}{\textit{Transformer-based indices}} \\
index\_stat            & 0.610 & 0.611 & 0.622 \\
index\_pos\_neg        & \textbf{0.627} & 0.630 & \textbf{0.638} \\
index\_pos\_neg\_max   & \textbf{0.627} & 0.630 & \textbf{0.638} \\
index\_ICI             & \textbf{0.627} & 0.630 & \textbf{0.638} \\
\midrule
\multicolumn{4}{l}{\textit{Vocabulary-based indices}} \\
index\_barbaglia       & 0.322 & 0.333 & 0.162 \\
index\_shapiro         & 0.322 & 0.333 & 0.162 \\
\bottomrule
\end{tabular}

\vspace{0.2cm}
\begin{minipage}{0.92\linewidth}
\footnotesize
\textit{Notes:} Models are trained using a temporal train--test split. Hyperparameters are selected via three-fold time-series cross-validation using macro-F1 as the optimization criterion. Reported values refer to performance on the hold-out test sample.
\end{minipage}
\end{table}
Although the machine-learning classifiers are trained directly on our manually labeled dataset, they only marginally outperform the transformer-based sentiment indices on the hold-out sample. This result is noteworthy because the indices are constructed from FinBERT sentiment scores rather than being optimized for our specific classification task. The limited performance gap suggests that the information captured by the FinBERT-based sentiment indices is already sufficient to approximate the overall mood expressed in the articles, leaving little additional predictive signal for more flexible classifiers to exploit. At the same time, these findings point to a promising direction for future research. Rather than relying on a general-purpose financial language model, it may be possible to develop a transformer specifically trained on our corpus, using document-level sentiment annotations instead of sentence-level predictions. Such a model could produce a document-level mood index that more accurately captures the contextual and narrative aspects of newspaper articles, potentially improving upon the current FinBERT-based index.

\paragraph{Out-of-Sample Regression Performance}

Table \ref{tab:oos_regression} reports the results of the out-of-sample regression performance. In general, all machine-learning models trained on transformer embeddings exhibit strong predictive performance on the hold-out test sample, explaining approximately 68--69\% of the variation in human sentiment ratings. The Random Forest achieves the highest out-of-sample fit ($R^2 = 0.688$), followed closely by Ridge regression with PCA ($R^2 = 0.685$) and Support Vector Regression ($R^2 = 0.679$). The similarity of these results suggests that the predictive signal contained in the transformer embeddings is robust across alternative modeling approaches.

Comparison with sentiment indices yields two additional insights. First, the transformer-based sentiment index (\texttt{index\_stat}) retains substantial predictive pow\-er out of sample ($R^2 = 0.632$), confirming that a single sentiment score can summarize a large share of the information contained in the full embedding representation. Second, vocabulary-based measures perform considerably worse, with out-of-sample $R^2$ values of 0.295 for the Shapiro index and 0.183 for the Barbaglia index. These results indicate that contextual language representations capture dimensions of sentiment that are largely missed by traditional vocabulary-based approaches.

Taken together, the out-of-sample evidence reinforces the findings of the correlation, classification, and in-sample regression analyses. Transformer-based methods consistently provide the closest approximation to human sentiment judgments and maintain their advantage when evaluated on previously unseen articles.

\paragraph{Model selection.}

Hyperparameters were selected using three-fold time-series cross-validation on the training sample, with the coefficient of determination ($R^2$) as the optimization criterion. For the Random Forest regressor, the search grid included the number of trees  $\in {100,200}$), maximum depth  $\in {3,4,5}$), minimum split size  $\in {2,5}$), and minimum leaf size . The selected specification used 200 trees, a maximum depth of 5, a minimum split size of 5, and a minimum leaf size of 2. For Ridge regression and Support Vector Regression (SVR), transformer embeddings were standardized and reduced using Principal Component Analysis (PCA). The optimal Ridge specification retained 10 principal components with a regularization parameter of $\alpha = 0.1$, while the best-performing SVR retained 10 principal components and employed a linear kernel with $C=10$ and $\gamma=\texttt{scale}$. Cross-validation scores were remarkably similar across models, ranging from $R^2=0.663$ for Random Forest to $R^2=0.668$ for SVR, suggesting that predictive performance is driven primarily by the information contained in the embeddings rather than by the specific regression algorithm.

\begin{table}[htbp]
\centering
\caption{Out-of-sample regression performance}
\label{tab:oos_regression}
\begin{tabular}{lccc}
\toprule
Method & $R^2$ & RMSE & MAE \\
\midrule
\multicolumn{4}{l}{\textit{Machine-learning models}} \\
Random Forest & \textbf{0.688} & \textbf{0.231} & 0.193 \\
Ridge + PCA   & 0.685 & 0.232 & 0.192 \\
SVR + PCA     & 0.679 & 0.234 & \textbf{0.191} \\
\midrule
\multicolumn{4}{l}{\textit{Transformer-based index}} \\
index\_stat   & 0.632 & 0.251 & 0.209 \\
\midrule
\multicolumn{4}{l}{\textit{Vocabulary-based indices}} \\
index\_shapiro   & 0.295 & 0.347 & 0.299 \\
index\_barbaglia & 0.183 & 0.374 & 0.324 \\
\bottomrule
\end{tabular}

\vspace{0.2cm}
\begin{minipage}{0.92\linewidth}
\footnotesize
\textit{Notes:} Models are estimated on a training sample and evaluated on a temporally separated hold-out test sample. Hyperparameters are selected via three-fold time-series cross-validation. Reported metrics include the coefficient of determination ($R^2$), root mean squared error (RMSE), and mean absolute error (MAE).
\end{minipage}
\end{table}

\section{Concluding Remarks}
\label{sec:conclusion}

In this paper, we revisit the construction of daily news-based mood indices by replacing traditional dictionary-based sentiment scoring with a context-sensitive transformer approach while preserving a transparent and interpretable aggregation framework. Using a corpus of more than 140,000 financial news articles from Dow Jones Factiva, we classify sentiment at the sentence level with FinBERT, construct alternative article-level sentiment measures that differ in their treatment of neutrality, their handling of sentiment imbalance between positive and negative content, and whether they incorporate the positional importance of sentiment-bearing sentences within the document and aggregate these into daily mood indices after controlling for publisher and calendar effects. This framework allows a direct comparison between transformer-based and vocabulary-based sentiment measures under a common empirical design.

Our results show that the method used to extract sentiment has important implications for the resulting mood indices. Transformer-based measures generate sentiment distributions that are more dispersed and more polarized than vocabulary-based benchmarks, which are concentrated around zero and therefore produce smoother, lower-amplitude signals. Despite differences in article-level aggregation, the alternative transformer-based indices display strong contemporaneous co-movement, indicating that the underlying daily sentiment dynamics are robust to the specific aggregation rule. Vocabulary-based indices capture similar broad movements over time but substantially compress sentiment variation, especially around the neutral category.

The analysis also highlights the importance of accounting for calendar effects when constructing daily mood indicators. Although daily sentiment exhibits only limited serial dependence beyond the first lag, both the level and dispersion of sentiment vary systematically across weekdays. Ignoring these regularities may therefore introduce unnecessary variation into high-frequency sentiment measures.

To assess which approach better reflects perceived sentiment, we validate the indices against a benchmark constructed from an incentivized human annotation exercise conducted on Prolific. Across correlations, regression models, and both in-sample and out-of-sample classification exercises, transformer-based measures consistently outperform vocabulary-based alternatives and provide a substantially closer approximation to human sentiment judgments. The out-of-sample analysis is particularly informative: machine-learning models trained directly on our manually labeled data improve only marginally upon the best transformer-based indices. This finding suggests that the information extracted by FinBERT already captures most of the signal required to reproduce human assessments of financial-news sentiment, leaving relatively limited scope for additional gains from downstream predictive models.

Methodologically, the paper demonstrates that sentiment measurement depends not only on the underlying language model but also on the aggregation strategy used to transform sentence-level predictions into document- and day-level indices. By systematically comparing alternative aggregation schemes within a common framework, we provide a transparent methodology that can be readily adapted to different datasets, languages, and macro-finance applications.

Several avenues for future research naturally emerge from our findings. First, while the results indicate that a pre-trained financial language model such as FinBERT provides an excellent basis for measuring news sentiment, future work could develop a transformer model trained directly on our annotated corpus. In particular, training the model using document-level human annotations rather than aggregating sentence-level predictions would allow it to learn sentiment from the complete narrative and contextual structure of financial news articles. Such an approach could produce a document-level mood index that more closely reflects human perceptions of economic news while remaining suitable for high-frequency monitoring. Second, future research should explore the broader economic applications of the proposed indices, including their ability to explain and forecast financial-market dynamics, macroeconomic expectations, and business-cycle fluctuations, extending recent work such as \cite{borgioli2024financial}.

Overall, the evidence suggests that transformer-based sentiment analysis represents more than a robustness check against dictionary-based methods. By exploiting contextual information, transformer models recover aspects of sentiment that vocabulary-based approaches systematically smooth out while maintaining an aggregation framework that is transparent, interpretable, and easy to replicate. Transformer-based daily mood indices therefore provide a more accurate and informative measure of financial-news sentiment and offer a promising tool for real-time monitoring and empirical research in macroeconomics and finance.

\newpage
\appendix

\setcounter{section}{0}
\setcounter{equation}{0}
\setcounter{table}{0}
\setcounter{figure}{0}

\numberwithin{equation}{section}
\counterwithin{table}{section}
\counterwithin{figure}{section}
\section{Construction of Document-Level Sentiment Measures}
\label{app:sentiment_indices}

We begin from sentence-level sentiment predictions generated by a multiclass classifier. For each sentence $s$, the model outputs log-probabilities over three sentiment classes—positive, negative, and neutral—denoted by
\[
\ell_s = (\ell_{s,\text{pos}}, \ell_{s,\text{neg}}, \ell_{s,\text{neu}}).
\]
The predicted sentiment label is assigned using the $\arg\max$ rule. Each sentence-level prediction is then mapped into a one-hot encoded vector in $\mathbb{R}^3$, where positive, negative, and neutral labels correspond respectively to $[1,0,0]$, $[0,1,0]$, and $[0,0,1]$. For a given document, sentence-level one-hot vectors are summed across all sentences, yielding document-level counts
\[
(pos,\, neg,\, neu) = \sum_s \text{one\_hot}(y_s),
\]
which represent the number of positive, negative, and neutral sentences in the text.

\paragraph{Baseline sentiment index.}
The first index, denoted \texttt{index\_stat}, measures net sentiment per sentence while explicitly accounting for neutral content. It is defined as
\[
\text{index\_stat} = \frac{pos - neg}{pos + neg + neu},
\]
and takes values in the interval $[-1,1]$. This index captures average sentiment intensity at the sentence level, with neutral sentences diluting polarity. As a result, it is stable and length-aware, but may understate sentiment when documents contain a large fraction of neutral sentences.

\paragraph{Early-sentiment indices.}
To account for the possibility that early content disproportionately shapes readers' perceptions, we construct a family of early-sentiment indices, denoted \texttt{index\_stat\_n\_fun($n$)}. For each document, we retain only the first $n$ sentences and compute the average one-hot prediction
\[
p = (p_{\text{pos}},\, p_{\text{neg}},\, p_{\text{neu}}).
\]
The early-sentiment index is then defined as $p_{\text{pos}} - p_{\text{neg}}$. This measure captures first-impression sentiment but discards later content and is sensitive to sentence ordering.

\paragraph{Polarity-conditional index.}
The second main index, \texttt{index\_pos\_neg}, focuses exclusively on polarized content by ignoring neutral sentences. It is defined as
\[
\text{index\_pos\_neg} =
\begin{cases}
0, & \text{if } pos = neg, \\[4pt]
\dfrac{pos - neg}{pos + neg}, & \text{otherwise},
\end{cases}
\]
and lies in $[-1,1]$. This index captures net polarity conditional on the presence of non-neutral sentences. While it emphasizes clearly polarized content, it can be unstable when the number of positive and negative sentences is small.

\paragraph{Dominance-based index.}
To measure the dominance of the prevailing sentiment, we define \texttt{index\_pos\_neg\_max} as
\[
\text{index\_pos\_neg\_max} =
\begin{cases}
0, & \text{if } pos = neg, \\[4pt]
\dfrac{pos - neg}{\max(pos,\, neg)}, & \text{otherwise}.
\end{cases}
\]
This index also lies in $[-1,1]$ and highlights how strongly the majority sentiment exceeds the minority. However, it is discontinuous at ties and discards information about the overall volume of sentiment-bearing sentences.

\paragraph{Tanh-calibrated margin index.}
To smooth extreme values and reduce sensitivity to large sentiment imbalances, we construct a tanh-calibrated index:
\[
\text{index\_pos\_neg\_tanh}
= \tanh\!\left(\frac{pos - neg}{\max(neu,\, 1)}\right).
\]
This specification normalizes the sentiment margin by the amount of neutral content, thereby penalizing overconfidence in documents with substantial neutral language. The lower bound on the denominator prevents instability when $neu$ is close to zero.

\paragraph{Investor Confidence Index (ICI).}
As an alternative aggregation measure, we adopt an Investor Confidence Index (ICI) inspired by \cite{antweiler2004all, liu2025emotion}, defined as the log ratio of positive to negative sentences:
\[
\text{ICI}=\ln\!\left(\frac{1+pos}{1+neg}\right).
\]
This transformation yields a signed indicator that increases as positive sentences become more prevalent and decreases as negative sentences dominate. By construction, \(\text{ICI}=0\) when \(pos=neg\); \(\text{ICI}>0\) indicates relatively more positive tone, whereas \(\text{ICI}<0\) indicates relatively more negative tone. The additive constants ensure numerical stability when either \(pos=0\) or \(neg=0\).

Because the index is the logarithm of a positive ratio, it is \emph{not} bounded in \([-1,1]\). Its support is
\[
\text{ICI}\in(-\infty,+\infty).
\]
In practice, the empirical range is much narrower because sentence counts are finite. The logarithmic transformation also emphasizes \emph{relative} differences between positive and negative counts (i.e., ratios), thereby compressing extreme values compared with linear count-based measures.

\paragraph{Illustrative example.}

The different document-level indices capture distinct dimensions of sentiment because they differ in how they treat \emph{neutral content}, \emph{sentiment imbalance}, and \emph{scale}. The baseline index (\texttt{index\_stat}) measures net sentiment relative to total sentences and is therefore \emph{length-aware} and strongly diluted by neutral content, making it conservative and stable for information-heavy articles. By contrast, \texttt{index\_pos\_neg} and \texttt{index\_pos\_neg\_max} focus only on polarized sentences: the former measures \emph{net polarity conditional on non-neutral content}, while the latter emphasizes \emph{dominance} of the majority sentiment over the minority. The tanh-based index (\texttt{index\_pos\_neg\_tanh}) combines both ideas by smoothing extreme margins and scaling the sentiment gap by neutral content, thus penalizing overconfidence when articles are mostly descriptive. Finally, the Investor Confidence Index (\texttt{ICI}) summarizes sentiment through a log-ratio of positive to negative counts; it is unbounded and depends on the \emph{relative} balance between positive and negative tone rather than on total document length. In practice, these indices are complementary: some are better suited for capturing broad document tone (\texttt{index\_stat}), others for extracting pure polarity from opinionated text (\texttt{index\_pos\_neg}, \texttt{index\_pos\_neg\_max}), and others for robust nonlinear scaling (\texttt{index\_pos\_neg\_tanh}, \texttt{ICI}).

To illustrate the differences across indices, consider a document with \(20\) sentences, of which \(pos=8\) are positive, \(neg=5\) are negative, and \(neu=7\) are neutral. The resulting document-level indices are
\[
\text{index\_stat}=0.15,\qquad
\text{index\_pos\_neg}\approx 0.23,\qquad
\text{index\_pos\_neg\_max}=0.375,
\]
\[
\text{index\_pos\_neg\_tanh}
\approx 0.40,\qquad
\text{ICI}\approx 0.41.
\]
For the early-sentiment index, if the first \(n=5\) sentences contain \(2\) positive, \(1\) negative, and \(2\) neutral labels, then \(\texttt{index\_stat\_n\_fun(5)}=p_{\text{pos}}-p_{\text{neg}}=2/5-1/5=0.20\).

Now consider a second document with the \emph{same sentiment margin} (\(pos-neg=3\)) but much more neutral content, e.g., \(pos=8\), \(neg=5\), \(neu=27\) (40 sentences in total). Then
\[
\text{index\_stat}\approx 0.23,\qquad
\text{index\_pos\_neg\_max}=0.375,
\]
\[
\text{index\_pos\_neg\_tanh}\approx 0.11,\qquad
\text{ICI}=\approx 0.41.
\]
This comparison shows that \texttt{index\_stat} and \texttt{index\_pos\_neg\_tanh} are sensitive to neutral-content dilution, whereas \texttt{index\_pos\_neg}, \texttt{index\_pos\_neg\_max}, and \texttt{ICI} depend only on the positive--negative balance.

\begin{table}[!htbp]
\centering
\footnotesize
\caption{Summary of document-level sentiment aggregation methods}
\label{tab:index_summary}
\renewcommand{\arraystretch}{1.12}
\setlength{\tabcolsep}{4pt}
\begin{tabular}{p{3.4cm}p{2.7cm}p{1.7cm}p{3.0cm}p{4.5cm}}
\toprule
\textbf{Index} & \textbf{What it captures} & \textbf{Range} & \textbf{Neutral sentences} & \textbf{Main trade-off} \\
\midrule
\texttt{index\_stat} &
Overall net sentiment per sentence &
$[-1,1]$ &
Included (dilution) &
Stable and length-aware, but conservative in neutral-heavy texts \\

\texttt{index\_stat\_($n$)} &
Early-sentence sentiment (first impression) &
$[-1,1]$ &
Included in first-$n$ average &
Captures opening tone, but ignores later content and depends on sentence order \\

\texttt{index\_pos\_neg} &
Net polarity among polarized sentences &
$[-1,1]$ &
Ignored &
Highlights polarity, but can be unstable when few polarized sentences are present \\

\texttt{index\_pos\_neg\_max} &
Dominance of majority sentiment &
$[-1,1]$ &
Ignored &
Emphasizes dominance, but ignores sentiment volume and is tie-sensitive \\

\texttt{index\_pos\_neg\_tanh} &
Smoothed polarity scaled by neutral content &
$[-1,1]$ &
Used in denominator (\(\max(neu,1)\)) &
Penalizes overconfidence and compresses extremes, but is less directly interpretable \\

\texttt{ICI} &
Log-ratio of positive vs.\ negative sentences &
$(-\infty,+\infty)$ &
Ignored &
Captures relative balance and compresses extremes, but is unbounded \\
\bottomrule
\end{tabular}
\end{table}

\section{Serial Dependence and Day-of-Week Effects}
\label{app:serial_weekday}
Table~\ref{tab:acf_raw_wd} indicates that daily sentiment indices exhibit only modest short-run persistence and limited autocorrelation beyond the first lag, both before and after weekday adjustment. A useful distinction emerges between transformer-based and vocabulary-based measures. In the raw series (Panel~a), the vocabulary-based indices (\texttt{index\_barbaglia} and \texttt{index\_shapiro}) show statistically significant positive autocorrelation at lag~1 (both around \(0.16\)), whereas most transformer-based indices have smaller and generally insignificant lag-1 coefficients (typically around \(0.09\)–\(0.13\), with \texttt{index\_ici} as an exception at \(0.131\)). This pattern is consistent with the smoother, more compressed dynamics of vocabulary-based indices. After removing weekday effects (Panel~b), the overall picture remains similar: lag-1 persistence is still modest, most intermediate lags remain close to zero, and there is little evidence of sustained serial dependence. At the same time, several indices---including both vocabulary-based measures and some transformer-based variants---display significant negative autocorrelation at longer horizons (especially lag~14), suggesting mild mean reversion or periodic adjustment rather than persistent momentum. Overall, the table shows that the main signal in daily mood indices is not strong serial persistence, and that vocabulary-based indices differ primarily by exhibiting slightly smoother short-run dynamics than the transformer-based measures.

\input{acf_table}

\subsection{Serial Dependence and Day-of-Week Effects}

\begin{figure}[h!]
\centering
\includegraphics[width=1.1\linewidth]{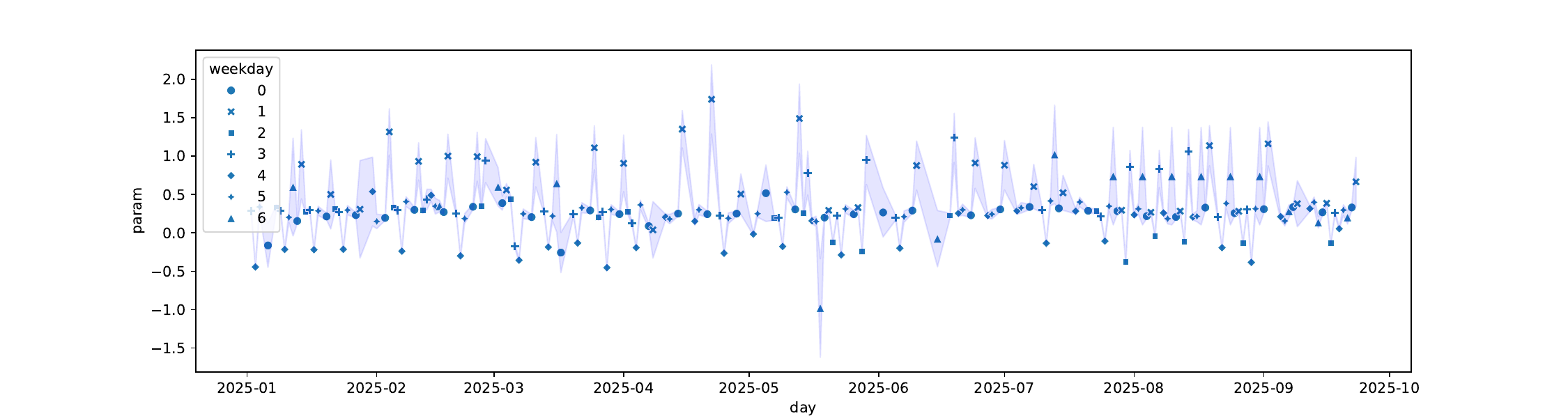}
\caption{Day-of-week patterns in daily sentiment (0 = Monday, 6 = Sunday). The figure shows that Tuesday exhibits the highest average sentiment level over time.}
\label{fig:weekday_patterns}
\end{figure}

As a robustness check, we examine serial dependence in the daily sentiment
measures before constructing moving-average indices. For each index, we compute
autocorrelations between the daily series and its $k$-day lag, for
$k=1,\ldots,14$. Because weekday patterns may mechanically generate apparent
persistence, especially at weekly horizons, we also residualize each daily series with respect to weekday fixed effects and recompute the autocorrelations on the residuals. Finally, we compute lag-$k$ correlations within weekday subsamples to assess whether weekly dependence is concentrated on specific calendar days.
Table~\ref{tab:acf_raw_wd} shows that the daily sentiment indices exhibit only
modest short-run persistence and little autocorrelation beyond the first lag, both before and after weekday adjustment. Vocabulary-based indices display somewhat stronger lag-1 autocorrelation than most transformer-based indices, consistent with their smoother and more compressed dynamics. After removing weekday effects, lag-1 persistence remains limited, intermediate lags are close to zero, and some indices display negative autocorrelation at longer horizons, especially around lag~14. Overall, the evidence does not indicate strong serial dependence in daily mood. Instead, part of the apparent weekly dependence appears to reflect
calendar-specific variation.
Table~\ref{tab:weekday_stats} reports summary statistics of the daily sentiment index aggregated by weekday, showing clear heterogeneity in sentiment levels across the week. Tuesday exhibits the highest average sentiment (mean = 0.744), together with relatively high dispersion, while Friday is the weakest day on average (mean = -0.025), suggesting slightly negative or neutral tone at the end of the working week. Midweek days show mixed patterns, with Wednesday displaying the lowest central tendency and occasional negative values, whereas Thursday shows a rebound in sentiment levels. Weekend days are generally more stable, with Saturday showing relatively low variance and mildly positive sentiment, while Sunday exhibits higher mean sentiment but also greater variability, indicating occasional extreme observations. Figure \ref{fig:weekday_patterns} of the sentiment index over time further shows that newspaper mood is highly dynamic, with clear short-run fluctuations and periods of elevated or reduced sentiment rather than a smooth trend, suggesting that sentiment is primarily driven by time-varying news shocks rather than long-run deterministic movements.

Table~\ref{tab:weekday_drivers} reports the correlation structure of all sentiment indices across weekdays at short-run (lag 1) and medium-run (lag 14) horizons. The results indicate clear evidence of short-run persistence, particularly on Tuesdays, where most indices exhibit positive and statistically significant lag-1 correlations, suggesting that sentiment shocks tend to persist within the weekly cycle. By contrast, Fridays show weaker and in some cases negative short-run persistence, consistent with faster mean reversion toward the end of the week. At the 14-day horizon, the pattern becomes more heterogeneous, with several indices displaying reversal effects, especially on Tuesdays and Fridays, where correlations turn negative or lose significance. Overall, the evidence points to a combination of short-run persistence and medium-run reversion, indicating that sentiment dynamics exhibit both momentum and cyclical adjustment depending on the weekday and the time horizon.

\begin{table}[h!]
\centering
\scriptsize
\caption{Summary statistics of daily sentiment by weekday}
\label{tab:weekday_stats}
\begin{tabular}{lrrrrrrrr}
\toprule
Weekday & count & mean & std & min & 25\% & 50\% & 75\% & max \\
\midrule
Monday    & 37 & 0.245 & 0.132 & -0.257 & 0.217 & 0.265 & 0.305 & 0.516 \\
Tuesday   & 32 & 0.744 & 0.418 & 0.039 & 0.367 & 0.770 & 0.994 & 1.739 \\
Wednesday & 20 & 0.129 & 0.236 & -0.378 & -0.116 & 0.240 & 0.294 & 0.434 \\
Thursday  & 28 & 0.416 & 0.337 & -0.174 & 0.226 & 0.281 & 0.522 & 1.238 \\
Friday    & 35 & -0.025 & 0.273 & -0.453 & -0.218 & -0.133 & 0.217 & 0.538 \\
Saturday  & 35 & 0.288 & 0.089 & 0.148 & 0.213 & 0.300 & 0.342 & 0.528 \\
Sunday    & 16 & 0.450 & 0.478 & -0.979 & 0.262 & 0.623 & 0.739 & 1.021 \\
\bottomrule
\end{tabular}
\end{table}

\input{weekday_drivers_table}

\subsection{Literature Review: Summary of Existing Evidence}
Table~\ref{tab:lit_review_readable} provides a structured overview of the existing literature on sentiment measurement in news and macro-finance, summarizing the scope, methodologies, and main findings of key contributions.
\begin{landscape}
\begin{table}[p]
\centering
\footnotesize
\renewcommand{\arraystretch}{1}
\setlength{\tabcolsep}{4pt}

\caption{Literature Review in a Table}
\label{tab:lit_review_readable}

\begin{tabularx}{\linewidth}{@{}p{3.4cm}p{1.5cm}p{3.6cm}X@{}}
\toprule
\textbf{Author(s) \& Year} & \textbf{Scope} & \textbf{Methodology} & \textbf{Key Findings} \\
\midrule
\cite{aprigliano2023power} & Italy & Italian Dict + Valence Shifters &
Monthly BMA and weekly trackers improve GDP nowcasts, with larger gains in recessions. \\

\cite{arcin2025constructing} & Philippines & vocabularys vs.\ FinBERT + NMF &
A FinBERT-based News Sentiment Index (NSI) captures domestic turning points effectively. \\

\cite{ashwin2024nowcasting} & Euro Area & Translation + Ridge/NN &
News sentiment adds most information early in the quarter, when soft data are limited. \\

\cite{barbaglia2024forecasting} & Europe & FiGAS + Neural Translation &
News indicators predict activity across major EU economies and remain robust in real-time settings. \\

\cite{barbaglia2025sentiment} & US / UK & Economic Lexicon (EL) &
Human-annotated EL scores outperform finance/accounting lexicons in macro applications. \\

\cite{aguilar2021can} & Spain & DENSI (Directional) &
A daily index detects COVID-19 regime shifts faster than traditional confidence surveys. \\

\cite{gross2024learning} & Global & Supervised Autoencoders &
Nonlinear latent factors from news analytics outperform linear PCA and PLS benchmarks. \\

\cite{rambaccussing2020forecasting} & UK & SVM + Keyword Indices &
Text-implied sentiment improves forecasts of UK unemployment and output growth. \\

\cite{lagerborg2023sentimental} & US & IV (Mass Shootings) + SVAR &
Identifies autonomous sentiment shocks associated with hump-shaped declines in industrial production. \\

\cite{lukauskas2022economic} & Lithuania & Transformers + Clustering &
AI-derived negative news sentiment reduces errors in forecasting economic activity. \\

\cite{magro2025can} & Euro Area & Native-Language Dictionaries &
Avoiding translation lowers resource costs while preserving nowcast accuracy. \\

\cite{ravi2025large} & Financial & LLMs (Llama3, RoBERTa) &
LLM-based sentiment (e.g., Llama3) achieves high classification accuracy and tracks sentiment-driven moves. \\

\cite{seo2024measuring} & S.\ Korea & Transformer Encoder (KoNS) &
Automated NSI signals inflection points 1--2 months ahead of official statistics. \\

\cite{shapiro2022measuring} & US & News PMI model &
PMI-weighted text sentiment behaves similarly to survey-based measures in macro models. \\

\cite{jiang2023financial} & US stocks & FinBERT + LSTM &
Combining FinBERT text signals with numeric inputs via LSTM improves price-movement prediction. \\

\cite{zhang2025interpretable} & Global FX & FinBERT + XGBoost + SHAP &
Interpretable ML delivers strong cost-adjusted performance in FX and bond markets (Sharpe $>4$). \\

\cite{chen2025structural} & Reviews & STS (Structural Topic/Sent) &
A structural topic--sentiment model links document covariates to topic prevalence and discourse slant. \\

\cite{de2025enhancing} & Euro Area & ChatGPT (Zero-shot) &
Activity scores from PMI releases improve GDP nowcasts by about 20\% versus hard-data benchmarks. \\

\cite{kalamara2022making} & UK & Dictionaries vs.\ ML (NN) &
Dictionary indices are useful baselines, but neural models yield larger marginal forecast gains. \\
\bottomrule
\end{tabularx}
\end{table}
\end{landscape}

\printbibliography[notkeyword=OWN]

\end{document}

%% file: acf_table.tex
\begin{sidewaystable}[!htbp]
\centering
\caption{Autocorrelation (lags 1--14): raw vs.\ weekday-adjusted daily indices}
\label{tab:acf_raw_wd}
\scriptsize
\setlength{\tabcolsep}{2pt}
\renewcommand{\arraystretch}{0.5}
\begin{tabular}{l*{14}{c}}
\toprule
 & (1) & (2) & (3) & (4) & (5) & (6) & (7) & (8) & (9) & (10) & (11) & (12) & (13) & (14) \\\\
\midrule
\multicolumn{15}{l}{\textit{(a) Raw}} \\\\
index\_barbaglia & 0.165** & 0.022 & 0.111 & -0.032 & 0.019 & 0.019 & -0.008 & 0.012 & -0.040 & 0.009 & -0.004 & -0.072 & -0.019 & -0.129* \\\\
  & (0.020) & (0.769) & (0.123) & (0.655) & (0.793) & (0.796) & (0.912) & (0.865) & (0.587) & (0.902) & (0.956) & (0.338) & (0.797) & (0.065) \\\\
index\_ici & 0.131* & -0.050 & -0.067 & -0.003 & 0.033 & 0.014 & 0.081 & -0.027 & -0.121 & -0.052 & -0.009 & -0.069 & -0.069 & -0.023 \\\\
  & (0.066) & (0.504) & (0.350) & (0.961) & (0.652) & (0.846) & (0.246) & (0.717) & (0.102) & (0.475) & (0.905) & (0.358) & (0.347) & (0.747) \\\\
index\_pos\_neg & 0.101 & -0.074 & -0.052 & -0.023 & 0.019 & -0.001 & 0.101 & 0.002 & -0.119 & -0.039 & -0.012 & -0.059 & -0.080 & -0.014 \\\\
  & (0.156) & (0.316) & (0.471) & (0.750) & (0.795) & (0.988) & (0.148) & (0.982) & (0.106) & (0.591) & (0.871) & (0.433) & (0.275) & (0.843) \\\\
index\_pos\_neg\_max & 0.103 & -0.071 & -0.050 & -0.017 & 0.021 & -0.008 & 0.106 & -0.009 & -0.128* & -0.032 & -0.010 & -0.073 & -0.079 & -0.015 \\\\
  & (0.148) & (0.340) & (0.489) & (0.812) & (0.773) & (0.910) & (0.130) & (0.898) & (0.084) & (0.660) & (0.892) & (0.334) & (0.281) & (0.830) \\\\
index\_pos\_neg\_tanh & 0.090 & 0.083 & -0.112 & 0.027 & 0.062 & -0.025 & -0.012 & -0.044 & -0.060 & 0.011 & -0.005 & -0.014 & -0.054 & -0.196*** \\\\
  & (0.206) & (0.262) & (0.119) & (0.702) & (0.399) & (0.732) & (0.869) & (0.548) & (0.414) & (0.875) & (0.947) & (0.848) & (0.462) & (0.005) \\\\
index\_shapiro & 0.161** & 0.032 & 0.093 & -0.039 & 0.000 & 0.023 & 0.053 & 0.010 & -0.050 & -0.016 & -0.013 & -0.086 & -0.043 & -0.189*** \\\\
  & (0.024) & (0.668) & (0.197) & (0.587) & (0.995) & (0.749) & (0.450) & (0.894) & (0.499) & (0.823) & (0.853) & (0.252) & (0.562) & (0.007) \\\\
index\_stat\_10 & 0.102 & -0.027 & -0.020 & 0.015 & 0.035 & -0.033 & -0.048 & 0.008 & -0.107 & 0.036 & -0.033 & -0.103 & -0.082 & -0.054 \\\\
  & (0.151) & (0.716) & (0.780) & (0.836) & (0.635) & (0.650) & (0.495) & (0.912) & (0.147) & (0.622) & (0.647) & (0.172) & (0.267) & (0.442) \\\\
index\_stat\_30 & 0.096 & -0.005 & -0.059 & 0.016 & 0.057 & -0.009 & 0.024 & -0.034 & -0.082 & -0.005 & -0.024 & -0.031 & -0.071 & -0.145** \\\\
  & (0.180) & (0.948) & (0.416) & (0.824) & (0.444) & (0.901) & (0.728) & (0.640) & (0.266) & (0.942) & (0.737) & (0.677) & (0.333) & (0.039) \\\\
\addlinespace
\multicolumn{15}{l}{\textit{(b) Weekday-adjusted}} \\\\
index\_barbaglia & 0.147** & 0.025 & 0.142** & -0.001 & 0.022 & 0.009 & -0.076 & 0.006 & -0.029 & 0.039 & 0.023 & -0.071 & -0.017 & -0.209*** \\\\
  & (0.040) & (0.738) & (0.049) & (0.988) & (0.771) & (0.906) & (0.278) & (0.932) & (0.696) & (0.589) & (0.749) & (0.346) & (0.812) & (0.003) \\\\
index\_ici & 0.147** & -0.059 & -0.007 & 0.077 & 0.035 & -0.010 & -0.032 & -0.052 & -0.124* & 0.017 & 0.051 & -0.077 & -0.096 & -0.157** \\\\
  & (0.039) & (0.425) & (0.928) & (0.280) & (0.640) & (0.887) & (0.647) & (0.475) & (0.094) & (0.812) & (0.481) & (0.306) & (0.190) & (0.024) \\\\
index\_pos\_neg & 0.114 & -0.082 & 0.017 & 0.069 & 0.024 & -0.030 & -0.019 & -0.026 & -0.119 & 0.044 & 0.052 & -0.070 & -0.116 & -0.163** \\\\
  & (0.110) & (0.271) & (0.819) & (0.338) & (0.745) & (0.680) & (0.782) & (0.721) & (0.108) & (0.541) & (0.472) & (0.353) & (0.113) & (0.020) \\\\
index\_pos\_neg\_max & 0.115 & -0.074 & 0.021 & 0.076 & 0.028 & -0.038 & -0.012 & -0.039 & -0.129* & 0.052 & 0.055 & -0.084 & -0.116 & -0.162** \\\\
  & (0.107) & (0.317) & (0.773) & (0.289) & (0.703) & (0.600) & (0.863) & (0.598) & (0.081) & (0.471) & (0.447) & (0.263) & (0.115) & (0.020) \\\\
index\_pos\_neg\_tanh & 0.092 & 0.075 & -0.088 & 0.062 & 0.058 & -0.023 & -0.064 & -0.043 & -0.063 & 0.039 & 0.021 & -0.035 & -0.052 & -0.253*** \\\\
  & (0.196) & (0.310) & (0.223) & (0.385) & (0.436) & (0.748) & (0.360) & (0.555) & (0.397) & (0.593) & (0.776) & (0.644) & (0.478) & (0.000) \\\\
index\_shapiro & 0.143** & 0.035 & 0.132* & -0.003 & 0.007 & 0.008 & 0.002 & -0.000 & -0.033 & 0.017 & 0.017 & -0.086 & -0.051 & -0.243*** \\\\
  & (0.044) & (0.640) & (0.067) & (0.969) & (0.930) & (0.914) & (0.972) & (0.997) & (0.651) & (0.811) & (0.813) & (0.256) & (0.491) & (0.001) \\\\
index\_stat\_10 & 0.086 & -0.008 & -0.003 & 0.040 & 0.066 & -0.044 & -0.123* & -0.004 & -0.079 & 0.058 & -0.017 & -0.080 & -0.088 & -0.137* \\\\
  & (0.230) & (0.910) & (0.966) & (0.579) & (0.375) & (0.546) & (0.080) & (0.952) & (0.288) & (0.422) & (0.813) & (0.290) & (0.232) & (0.051) \\\\
index\_stat\_30 & 0.086 & -0.005 & -0.031 & 0.053 & 0.062 & -0.014 & -0.035 & -0.038 & -0.072 & 0.026 & 0.003 & -0.035 & -0.069 & -0.209*** \\\\
  & (0.226) & (0.949) & (0.669) & (0.456) & (0.403) & (0.849) & (0.615) & (0.601) & (0.328) & (0.718) & (0.970) & (0.639) & (0.348) & (0.003) \\\\
\bottomrule
\end{tabular}
\vspace{2mm}
\begin{minipage}{0.95\textwidth}\footnotesize
\textit{Notes:} Panel (a) reports raw autocorrelations of the daily index. Panel (b) reports autocorrelations of residuals after removing weekday fixed effects (regressing on $i.\text{weekday}$). Stars denote significance based on the correlation $p$-value (in parentheses): $^{*}p<0.10$, $^{**}p<0.05$, $^{***}p<0.01$. 
\end{minipage}
\end{sidewaystable}

%% file: weekday_drivers_table.tex
\pagestyle{empty}

\begin{table}[h]
\centering
\caption{Weekday drivers of dependence (lags 1 and 14)}
\label{tab:weekday_drivers}
\scriptsize
\setlength{\tabcolsep}{1pt}
\renewcommand{\arraystretch}{0.2}
\begin{tabular}{llc*{7}{c}}
\toprule
Index & Series & Lag & Mon & Tue & Wed & Thu & Fri & Sat & Sun \\\\
\midrule
index\_shapiro & raw & 1 & 0.191 & 0.538*** & 0.353 & -0.065 & -0.070 & 0.256* & 0.530** \\\\
 &  &  & (0.359) & (0.001) & (0.134) & (0.764) & (0.706) & (0.093) & (0.013) \\\\
\\[-3pt]
index\_shapiro & raw & 14 & -0.094 & -0.325* & 0.676** & 0.193 & -0.455*** & 0.034 & -0.161 \\\\
 &  &  & (0.538) & (0.055) & (0.042) & (0.365) & (0.005) & (0.828) & (0.508) \\\\
\\[-3pt]
index\_shapiro & wdresid & 1 & 0.191 & 0.538*** & 0.353 & -0.065 & -0.070 & 0.256* & 0.530** \\\\
 &  &  & (0.359) & (0.001) & (0.134) & (0.764) & (0.706) & (0.093) & (0.013) \\\\
\\[-3pt]
index\_shapiro & wdresid & 14 & -0.094 & -0.325* & 0.676** & 0.193 & -0.455*** & 0.034 & -0.161 \\\\
 &  &  & (0.538) & (0.055) & (0.042) & (0.365) & (0.005) & (0.828) & (0.508) \\\\
\\[-3pt]
\midrule
index\_barbaglia & raw & 1 & 0.206 & 0.424*** & 0.375 & 0.021 & -0.048 & 0.389** & 0.519** \\\\
 &  &  & (0.323) & (0.007) & (0.112) & (0.924) & (0.796) & (0.011) & (0.015) \\\\
\\[-3pt]
index\_barbaglia & raw & 14 & 0.019 & -0.319* & 0.590* & 0.204 & -0.492*** & 0.022 & -0.143 \\\\
 &  &  & (0.898) & (0.059) & (0.077) & (0.338) & (0.002) & (0.888) & (0.556) \\\\
\\[-3pt]
index\_barbaglia & wdresid & 1 & 0.206 & 0.424*** & 0.375 & 0.021 & -0.048 & 0.389** & 0.519** \\\\
 &  &  & (0.323) & (0.007) & (0.112) & (0.924) & (0.796) & (0.011) & (0.015) \\\\
\\[-3pt]
index\_barbaglia & wdresid & 14 & 0.019 & -0.319* & 0.590* & 0.204 & -0.492*** & 0.022 & -0.143 \\\\
 &  &  & (0.898) & (0.059) & (0.077) & (0.338) & (0.002) & (0.888) & (0.556) \\\\
\\[-3pt]
\midrule
\midrule
index\_ici & raw & 1 & -0.060 & 0.450*** & 0.532** & 0.010 & 0.128 & 0.020 & 0.364* \\\\
 &  &  & (0.775) & (0.004) & (0.024) & (0.964) & (0.492) & (0.896) & (0.088) \\\\
\\[-3pt]
index\_ici & raw & 14 & -0.001 & -0.406** & 0.362 & -0.294 & 0.240 & 0.096 & 0.023 \\\\
 &  &  & (0.996) & (0.016) & (0.277) & (0.168) & (0.139) & (0.542) & (0.923) \\\\
\\[-3pt]
index\_ici & wdresid & 1 & -0.060 & 0.450*** & 0.532** & 0.010 & 0.128 & 0.020 & 0.364* \\\\
 &  &  & (0.775) & (0.004) & (0.024) & (0.964) & (0.492) & (0.896) & (0.088) \\\\
\\[-3pt]
index\_ici & wdresid & 14 & -0.001 & -0.406** & 0.362 & -0.294 & 0.240 & 0.096 & 0.023 \\\\
 &  &  & (0.996) & (0.016) & (0.277) & (0.168) & (0.139) & (0.542) & (0.923) \\\\
\\[-3pt]
\midrule
index\_pos\_neg\_max & raw & 1 & -0.157 & 0.458*** & 0.491** & -0.088 & 0.070 & 0.012 & 0.290 \\\\
 &  &  & (0.450) & (0.003) & (0.037) & (0.687) & (0.707) & (0.935) & (0.174) \\\\
\\[-3pt]
index\_pos\_neg\_max & raw & 14 & 0.007 & -0.436*** & 0.284 & -0.065 & 0.304* & 0.067 & -0.067 \\\\
 &  &  & (0.966) & (0.010) & (0.395) & (0.761) & (0.061) & (0.670) & (0.784) \\\\
\\[-3pt]
index\_pos\_neg\_max & wdresid & 1 & -0.157 & 0.458*** & 0.491** & -0.088 & 0.070 & 0.012 & 0.290 \\\\
 &  &  & (0.450) & (0.003) & (0.037) & (0.687) & (0.707) & (0.935) & (0.174) \\\\
\\[-3pt]
index\_pos\_neg\_max & wdresid & 14 & 0.007 & -0.436*** & 0.284 & -0.065 & 0.304* & 0.067 & -0.067 \\\\
 &  &  & (0.966) & (0.010) & (0.395) & (0.761) & (0.061) & (0.670) & (0.784) \\\\
\\[-3pt]
\midrule

index\_stat\_30 & raw & 1 & 0.033 & 0.359** & 0.412* & -0.047 & -0.015 & 0.011 & 0.210 \\\\
 &  &  & (0.873) & (0.022) & (0.081) & (0.829) & (0.937) & (0.943) & (0.325) \\\\
\\[-3pt]
index\_stat\_30 & raw & 14 & 0.013 & -0.457*** & 0.387 & -0.360* & 0.216 & 0.073 & -0.110 \\\\
 &  &  & (0.932) & (0.007) & (0.246) & (0.091) & (0.183) & (0.645) & (0.650) \\\\
\\[-3pt]
index\_stat\_30 & wdresid & 1 & 0.033 & 0.359** & 0.412* & -0.047 & -0.015 & 0.011 & 0.210 \\\\
 &  &  & (0.873) & (0.022) & (0.081) & (0.829) & (0.937) & (0.943) & (0.325) \\\\
\\[-3pt]
index\_stat\_30 & wdresid & 14 & 0.013 & -0.457*** & 0.387 & -0.360* & 0.216 & 0.073 & -0.110 \\\\
 &  &  & (0.932) & (0.007) & (0.246) & (0.091) & (0.183) & (0.645) & (0.650) \\\\
\\[-3pt]
\bottomrule
\end{tabular}
\vspace{2mm}
\begin{minipage}{0.95\textwidth}\footnotesize
\textit{Notes:} Each cell reports the correlation between the current-day value and its lagged value within the indicated weekday subsample (Monday--Sunday), for lags \(k \in \{1,14\}\). Statistical significance is based on the correlation \(p\)-value and is denoted by \(^{*}p<0.10\), \(^{**}p<0.05\), and \(^{***}p<0.01\). The second line in each cell reports the corresponding \(p\)-value in parentheses.
\end{minipage}
\end{table}